\documentclass[11pt]{article}

\usepackage{slashed}

\newcommand{\K}{{\mathcal K}}
\newcommand{\F}{{\mathcal F}}
\newcommand{\W}{{\mathcal W}}

\newcommand{\eol}{\notag \\}

\newcommand{\bj}{{\bar j}}

\newcommand{\bphi}{{\bar \phi}}

\newcommand{\beq}{\begin{equation}}
\newcommand{\eeq}{\end{equation}}
\newcommand{\ul}{\underline}

\newcommand{\eps}{{\epsilon}}

\newcommand{\hc}{\mathrm{h.c.}}

\newcommand{\Tr}{\mathrm{Tr }}

\newcommand{\piecett}[4]
        {\left\{\begin{array}{cc}
                #1 & #2 \\
                #3 & #4
        \end{array} \right.}

\newcommand{\CD}{\mathcal{D}}
\newcommand{\BCD}{\mathcal{\bar D}}

\newcommand{\CP}{{\mathcal P}}
\newcommand{\ACP}{{\bar {\mathcal P}}}

\newcommand{\dalpha}{{\dot{\alpha}}}
\newcommand{\dbeta}{{\dot{\beta}}}
\newcommand{\dgamma}{{\dot{\gamma}}}

\newcommand{\dmu}{{\dot{\mu}}}

\newcommand{\dphi}{{\dot{\phi}}}
\newcommand{\btheta}{{\bar\theta}}

\newcommand{\chE}{\mathcal E}

\newcommand{\bPhi}{{\bar \Phi}}

\usepackage{amsmath}
\usepackage{amsfonts}
\usepackage{wasysym}
\usepackage{graphicx}

\numberwithin{equation}{section}

\begin{document}
\thispagestyle{empty}


\hfill *UCB-PTH-09/28

%
\hfill September 30, 2009

\addvspace{45pt}

\begin{center}

\Large{\textbf{Background field formalism for chiral matter and
gauge fields conformally coupled to supergravity}}
\\[35pt]
\large{Daniel Butter}
\\[10pt]
\textit{Department of Physics, University of California, Berkeley}
\\ \textit{and}
\\ \textit{Theoretical Physics Group, Lawrence Berkeley National Laboratory}
\\ \textit{Berkeley, CA 94720, USA}
\\[10pt] 
dbutter@berkeley.edu
\end{center}

\addvspace{35pt}

\begin{abstract}
\noindent
We expand the generic model involving chiral matter, super Yang-Mills
gauge fields, and supergravity to second order in the gravity and gauge
prepotentials in a manifestly covariant and conformal way. Such a class
of models includes conventional chiral matter coupled to supergravity
via a conformal compensator. This is a first step toward calculating
one-loop effects in supergravity in a way that does not require a
perturbative expansion in the inverse Planck scale or a recourse to
component level calculations to handle the coupling of the K\"ahler potential
to the gravity sector.
We also consider a more restrictive model involving a linear superfield
in the role of the conformal compensator and investigate the similarities
it has to the dual chiral model.

\end{abstract}

\setcounter{tocdepth}{2}
\newpage
\setcounter{page}{1}

\section{Introduction}
The background approach to quantization has a long pedigree in
superspace approaches to supergravity. The important work of Grisaru
and Siegel \cite{Grisaru:1981xm, Grisaru:1982zh}
(extended later by Grisaru and Zanon \cite{Grisaru:1983rg, Grisaru:1984ja, Grisaru:1984jc}
to include off-shell background fields)
showed how to expand old minimal Poincar\'e supergravity in terms of fundamental
quantum variations about a classical background, but they
restricted their consideration to old minimal supergravity alone.
This is difficult enough to do given the constrained supergeometry,
and its quantization requires the introduction of
not only Fadeev-Popov ghosts but also ghosts for ghosts, Nielsen-Kallosh
ghosts \cite{nkghost}, and ``hidden'' ghosts \cite{hidden_ghosts} which a casual application
of the Fadeev-Popov procedure might miss. The on-shell one-loop gauge-fixed quantum
Lagrangian was found which allows certain simple calculations as well as the
construction of covariant Feynman rules to handle more general
theories perturbatively. This story is by now textbook material \cite{superspace}.

However, the calculation of even one-loop effects
involving not only supergravity but also chiral matter and gauge
fields has to our knowledge never been comprehensively undertaken
in superspace. Part of this is undoubtedly the difficulty in dealing
with not only the constrained structure of supergravity in superspace but
also the Brans-Dicke coupling of chiral matter to the superspace
Einstein-Hilbert term. In a purely Poincar\'e approach, this last feature
requires either a component space Weyl rescaling \cite{wb} or the introduction
of $U(1)$ superspace and a superfield Weyl rescaling \cite{bgg}.
In this respect, it is almost more straightforward to work at the component level
and then to extract superspace results from the component ones.
A conformal approach at the superfield level seems a more feasible method,
and that is the approach we take here.

We have begun a program to attempt the calculation of one-loop corrections
to an arbitrary chiral model coupled to super Yang-Mills and supergravity
within superspace, thus maintaining manifest supersymmetry at all stages.
In order to deal ultimately with the conformal coupling of the canonical
K\"ahler potential in the Einstein-Hilbert term, we have shown how,
in a previous work, to extend the structure group of Poincar\'e superspace
to include the superconformal group \cite{Butter:2009cp}.
The new conformally covariant derivatives possess an algebra which is
identical to that of gauge theories: their curvatures are expressed
in terms of ``gaugino'' superfields $\W_\alpha$ and $\W^\dalpha$
valued in the superconformal group, which obey a generalized chirality
condition \eqref{eq_chiralW} as well as a Bianchi identity \eqref{eq_bianchiW}.
The selection of a number of curvature constraints eliminate most of
the these superfields, and the ones which remain may all be described
by the single chiral superfield $W_{\alpha \beta \gamma}$, the
chiral spinor field strength of conformal supergravity. The conformally
covariant derivatives and their curvatures all transform covariantly
under the superconformal algebra, which simplifies the calculation
of superscale transformations considerably.

Were it not for the constraints on the $\W_\alpha$, the structure
of the theory would be quite easy to solve. In analogy with
Yang-Mills, one would expect unconstrained prepotentials $V^A$,
one for each member of the superconformal algebra. The constraints
on the curvatures clearly must eliminate most of these prepotentials
since a large volume of literature (see for example the textbooks
\cite{superspace, Buchbinder:1998qv} as well as the original work \cite{Ogievetsky:1978mt})
shows that the fundamental quanta of old minimal Poincar\'e supergravity are the superfields
$H^M = (H^m, H^\mu, H_\dmu)$ and a chiral compensator $\sigma$,
with a gauge invariance allowing one to algebraically eliminate
$H^\mu$ and $H_\dmu$. We will not attempt to solve the constraints on
the full prepotentials here. Rather, as our interest is in performing one-loop
calculations in a classical background, we will focus on calculating
the allowed deformations of the prepotentials which preserve the
curvature constraints. The degrees of freedom must, of course, be the
same in either approach.

This paper is composed of three sections. In the first, we establish
that the theory, like Yang-Mills, is defined in terms of prepotentials.
We study arbitrary first order deformations of the prepotentials and solve
for the form that leave the constraints invariant to first order.
In the second section, we consider two physical actions, one involving
the arbitrary coupling of chiral superfields to supergravity and the
other involving the minimal linear compensator model with a K\"ahler
potential. We construct their first order variations in terms of their
fundamental quanta about a classical background and demonstrate that they
possess a common structure.
In the third section, we proceed to second order and present the second
order variation of the action for both models, which is sufficient (after
gauge fixing) for one-loop computations.

\section{Prepotential formulation of conformal superspace}
The algebra of the conformally covariant derivatives are \cite{Butter:2009cp}
\begin{gather}
\{\nabla_\alpha, \nabla_\beta\} = 0, \;\;\;
\{\nabla_\dalpha, \nabla_\dbeta\} = 0 \eol
\{\nabla_\alpha, \nabla_\dalpha\} = -2i \nabla_{\alpha \dalpha} \eol
\{\nabla_\beta, \nabla_{\alpha \dalpha}\} = -2i \eps_{\beta \alpha} \W_\dalpha, \;\;\;
\{\nabla_\dbeta, \nabla_{\alpha \dalpha}\} = -2i \eps_{\dbeta \dalpha} \W_\alpha \label{eq_galg}
\end{gather}
where $\W_\alpha$ are the ``gaugino superfields'' for the superconformal group.
These superfields are covariantly chiral in the sense that
\begin{align}\label{eq_chiralW}
\{\nabla_\dalpha, \W_\alpha\} = 0, \;\;\;
\{\nabla_\alpha, \W_\dalpha\} = 0
\end{align}
and obey the Bianchi identity
\begin{align}\label{eq_bianchiW}
\{\nabla^\alpha, \W_\alpha\} = \{\nabla_\dalpha, \W^\dalpha\}
\end{align}
The structure is clearly reminiscent of Yang-Mills, except for two
differences: the gauge generators $X_B$ do not commute with the
covariant derivatives ($[X_B, \nabla_A] \neq 0$), and most of
the $\W_\alpha$ are constrained to vanish.
The combination of the constraints and the Bianchi identities
then allow one to solve for the non-vanishing $\W_\alpha$
all in terms of the single chiral superfield $W_{\alpha \beta \gamma}$.

The structure of the covariant derivatives of conformal supergravity
allows a solution in terms of prepotentials that is identical in its
structure to that of gauge theories. For example, \eqref{eq_galg}
implies the existence of a chiral (+) and an antichiral (-) gauge where
\begin{align}
\nabla^\dalpha{}^{(+)} = \partial^\dalpha =  T \nabla^\dalpha T^{-1}, \;\;\;
\nabla_\alpha^{(-)} = \partial_\alpha = \bar T \nabla_\alpha \bar T^{-1}
\end{align}
where $T$ and $\bar T$ represent the superconformal gauge transformations
taking us from an arbitrary gauge to the two special ones.
Inverting these formulae gives
\begin{align}
\nabla_\alpha = \bar T^{-1} \partial_\alpha \bar T, \;\;\;
\nabla_\dalpha =  T^{-1} \partial_\dalpha T
\end{align}
which serve to encode the details of the connections in an arbitrary
gauge in terms of a complex gauge prepotential $T$.

It is clear that the special gauges $T$ and $\bar T$ are ill-defined
up to transformations of the form
\begin{align}
T \rightarrow C T, \;\;\;
\bar T \rightarrow \bar C \bar T
\end{align}
where $C$ is chiral ($[\partial_\dalpha, C] = 0$) and $\bar C$ is
antichiral ($[\partial_\alpha, \bar C] = 0$). In addition, they
transform under gauge transformations as
\begin{align}
T \rightarrow T G^{-1}, \;\;\;
\bar T \rightarrow \bar T G^{-1}
\end{align}
Putting these two transformations together gives a combined gauge/chiral
transformation of the form
\begin{align}
T \rightarrow C T G^{-1}, \;\;\;
\bar T \rightarrow \bar C \bar T G^{-1}
\end{align}

It is convenient to define the object $U \equiv \bar T T^{-1}$, which
represents the gauge transformation from the chiral to the antichiral
gauge. That is, $\nabla^{(-)}_A = U \nabla^{(+)}_A U^{-1}$.
Applying this formula and its inverse in the cases where the covariant
derivative is simple leads to
\begin{gather}
\nabla_\alpha^{(-)} = \partial_\alpha, \;\;\;
\nabla_\dalpha^{(-)} = U \partial_\dalpha U^{-1}  \eol
\nabla_\alpha^{(+)} = U^{-1} \partial_\alpha U, \;\;\;
\nabla_\dalpha^{(+)} = \partial_\dalpha
\end{gather}
$U$ is invariant under the full gauge transformations but transforms under
chiral gauge transformations as
\begin{align}
U \rightarrow \bar C U C^{-1}.
\end{align}

A (covariantly) chiral superfield $\Phi$ is a superfield constrained to obey
$\nabla_\dalpha \Phi = 0$. This is not in practice a difficult
constraint to satisfy. In the chiral gauge, we define the
conventionally chiral superfield $\phi$ by $\phi \equiv \Phi^{(+)}$.
The chirality condition is then simply the analytic statement that
$\phi = \phi(x,\theta)$ is independent of $\btheta$. In any other gauge, we have
\begin{align}
\Phi = T^{-1} \Phi^{(+)} = T^{-1} \phi
\end{align}
While $\Phi$ transforms under a gauge transformation as
$\Phi \rightarrow G \Phi$, the conventionally chiral $\phi$
transforms as $\phi \rightarrow C \phi$ where $C$ is the
chiral gauge transformation parameter.
One may make an analogous statement about antichiral superfields:
\begin{align}
\Phi^\dag = \bar T^{-1} \Phi^{\dag (-)} = \bar T^{-1} \bphi
\end{align}
Under a gauge transformation, $\Phi$ and $\Phi^\dag$ transform
covariantly while $\phi$ and $\bphi$ transform as
\begin{align}
\phi \rightarrow C \phi, \;\;\;
\bphi \rightarrow \bar C \bphi
\end{align}

The canonical kinetic action for $\Phi$ can be rewritten in
terms of the conventionally chiral superfields
\begin{align}
\int E\, \Phi^\dag \Phi = \int E\, (\bar T^{-1}\bphi) (T^{-1} \phi)
\end{align}
Since the action is gauge-invariant (provided $\Phi$ is of
scaling dimension $\Delta=1$), we may perform a
gauge transformation with parameter $G = \bar T$; this gives
\begin{align}
\int E\, \bphi (\bar T T^{-1} \phi) = \int E\, \bphi (U \phi)
\end{align}
The equality of the above two statements is formally equivalent to
$\bar T^T = \bar T^{-1}$ where transposition is understood as
moving the gauge generator off one term and onto another. (An
integration by parts, of course, has the same property.) One may use this
to adopt a notation where the kinetic term is written as
\begin{align}
\Phi^\dag \Phi = \bphi U \phi
\end{align}
where $U$ may be understood as acting either to the right (as $U$)
or to the left (as $U^{-1}$).

It is often useful to work in a Hermitian gauge. We denote such a gauge by $(0)$;
it is easily found by interpolating between the chiral and antichiral gauges:
\begin{gather}
\nabla_\alpha^{(0)} = U^{-1/2} \partial_\alpha U^{1/2}, \;\;\;
\nabla_\dalpha^{(0)} = U^{1/2} \partial_\dalpha U^{-1/2}
\end{gather}

We note that it is often useful to represent $U$ in an exponential form. We choose
to define the superfield $V^A$ by
\begin{align}
U =  \exp(-2 i V^A X_A)
\end{align}
Under this definition, $V^A$ is Hermitian and represents the superconformal
analogue of the gauge prepotential. If the constraints \eqref{eq_galg}
were the sole constraints on the geometry, the prepotentials $V^A$
would be unconstrained. However, certain of the gaugino superfields
$\W_\alpha$ are constrained to vanish, which serves to implicitly
define some of the $V^A$ in terms of the others. Experience in
Poincar\'e geometry tells us that $V^a$ is the unconstrained object
out of which the others are defined.\footnote{In the literature, $V^a$ is usually
replaced with $H^m$ and would be defined from the above with the coordinate
derivative $\partial_M$ replacing the covariant $\nabla_A$ in the
set of generators.} We will not be concerned, however, with presenting a full solution of the
constraints. Rather, as we are more concerned with one loop calculations
around a classical background, we will seek to construct the $V^A$ associated
with the quantum deformations themselves.

\subsection{Quantum deformations of conformal geometry}
The standard recipe for quantum calculations in supergravity involves splitting
the geometry into a background geometry and quantum fluctuations about that
background. Since the gauge connections are encoded in $T$ and $\bar T$ (and thereby in $U$),
splitting the former into a background and quantum contribution is accomplished
by doing the same with the latter. The method of splitting we will adopt is
\begin{align}
T \rightarrow T T_Q, \;\;\;
\bar T \rightarrow \bar T \bar T_Q
\end{align}
which corresponds to
\begin{align}
\nabla_\alpha \rightarrow \bar T_Q^{-1} \nabla_\alpha \bar T_Q, \;\;\;
\nabla_\dalpha \rightarrow T_Q^{-1} \nabla_\dalpha T_Q.
\end{align}
The new covariant derivatives can then be constructed perturbatively
out of the old ones. Similarly, chiral superfields transform under these
variations as
\begin{align}
\Phi \rightarrow T_Q^{-1} \Phi, \;\;\;
\bPhi \rightarrow \bar T_Q^{-1} \bPhi
\end{align}

The prepotentials transform under the combined chiral and supergauge transformations
as
\begin{align}
T T_Q \rightarrow C T T_Q G^{-1}, \;\;\;
\bar T \bar T_Q \rightarrow \bar C \bar T \bar T_Q G^{-1}.
\end{align}
Just as in the component case, the gauge transformation can be interpreted
as either a background or a quantum transformation.
As a background transformation, we take $T$ and $\bar T$ to transform as
\begin{align}
T \rightarrow C T G^{-1}, \;\;\;
\bar T \rightarrow \bar C \bar T G^{-1}.
\end{align}
and the quantum prepotentials to transform homogeneously
\begin{align}
T_Q \rightarrow G T_Q G^{-1}, \;\;\;
\bar T_Q \rightarrow G \bar T_Q G^{-1}
\end{align}
In practice, we will leave the background gauge unspecified; indeed, we will attempt
to maintain background gauge covariance at all times.

As a quantum transformation, $T$ is invariant and $T_Q$ transforms as
\begin{align}
T_Q \rightarrow C_Q T_Q G_Q^{-1}, \;\;\;
\bar T_Q \rightarrow \bar C_Q \bar T_Q G_Q^{-1}
\end{align}
where $C_Q \equiv T^{-1} C T$ and $\bar C_Q \equiv \bar T^{-1} \bar C \bar T$
are chiral and antichiral operators, obeying respectively
\begin{align}
0 = [\nabla_\alpha, \bar C_Q] = [\nabla^\dalpha, C_Q]
\end{align}
Henceforth, we will be concerned only with quantum
transformations. The supergauge freedom of $G_Q$ can be eliminated by choosing
to work in quantum chiral, antichiral, or Hermitian gauge.

We prefer to work in a gauge which maintains manifest Hermiticity at all times,
though it may occasionally be more cumbersome, so we choose the last of these gauges.
To go to quantum Hermitian gauge, one takes
$G_Q = \bar T_Q^{-1} U_Q^{1/2} = T_Q^{-1} U_Q^{-1/2}$ where $U_Q \equiv \bar T_Q T_Q^{-1}$.
This yields $T_Q = U_Q^{-1/2}$, $\bar T_Q = U_Q^{1/2}$, giving
\begin{align}
\nabla_\alpha' = U_Q^{-1/2} \nabla_\alpha U_Q^{+1/2}, \;\;\;
\nabla_\dalpha' = U_Q^{+1/2} \nabla_\dalpha U_Q^{-1/2}
\end{align}
for the covariant derivatives and
\begin{align}
\Phi' = U_Q^{1/2} \Phi, \;\;\;
\bPhi' = U_Q^{-1/2} \bPhi
\end{align}
for the chiral and antichiral superfields. The residual gauge transformation acts on $U_Q$ as
\begin{align}
U_Q \rightarrow \bar C_Q U_Q C_Q^{-1}
\end{align}

Quantum chiral gauge consists of making the quantum gauge choice
$T_Q = 1$, $\bar T_Q = U_Q$. In this approach, $\nabla_\dalpha$ remains
unchanged under quantum deformations of the geometry and so chiral superfields
remain unchanged. Quantum antichiral gauge is analogously constructued.

It is worth noting the relation between $U_Q$ and $U'$ in
background Hermitian gauge:
\begin{align}
U ' = \bar T' T'^{-1} = \bar T \bar T_Q T_Q^{-1} T^{-1} = \bar T U_Q T^{-1} = U^{1/2} U_Q U^{1/2}
\end{align}

\subsection{Conformally covariant quantum prepotentials}
The perturbative quantum prepotentials are the Hermitian superfields $V$
defined by\footnote{Notational consistency would demand that the $V$'s
be subscripted with $Q$'s to denote that they are quantum prepotentials.
Since we will never again mention the background prepotentials, it is easier
to suppress the $Q$ for a less cluttered notation.}
\begin{align}
U_Q = \exp\left(-2i V^B \nabla_B - 2i V^{\ul b} X_{\ul b} \right)
\end{align}
To maintain general covariance, we have chosen to parametrize the quantum
prepotentials in terms of the background covariant derivatives $\nabla_B$
rather than the coordinate derivatives. The factor of $-2$ is conventional
and the $i$ is so that the superfields $V^B$ have the obvious Hermiticity
conditions -- for example,
\begin{align}
(V^b)^\dag = V^b, \;\;\; (V^{\alpha})^\dag = V_\dalpha
\end{align}

These superfields are chosen to transform under the action of the group
generators as
\begin{align}
X_{\ul b} V^A = - V^C {f_{C \ul b}}^A
\end{align}
where $A$ and $C$ run over all indices and $f_{CB}{}^A$ are the structure
constants as defined in \cite{Butter:2009cp}. We thus have a conformally covariant
set of quantum prepotentials.

For the generators $D$ and $A$, the $V$'s transform contravariantly
as their index indicates. Thus $V^a$ (like ${e_m}^a$) has scaling
and $U(1)_R$ weights $(\Delta, w) = (-1,0)$,
$V^\alpha$ (like ${\psi_m}^\alpha$) has weights $(-1/2,+1)$,
but $V(K)^\alpha$ has weights $(+1/2,-1)$. For the Lorentz
generators, the $V$'s transform as their indices indicate.
Only special conformal transformation properties are not obvious.
Recall the action of $K$ on a group element $g = (\xi,\omega,\Lambda,w,\eps)$ is
\begin{gather}
K_B \xi^A = -\frac{1}{2} {{C_B}^A}_c \xi^c, \;\;\;
\frac{1}{2} (K_B \omega^{dc}) M_{cd} = -2 \xi^C M_{CB} \eol
K_B \Lambda = -2 (-)^B \xi_B , \;\;\;
K_B w = -3i \xi_B w(B) \eol
K_B \eps^A = -\lambda(A) \Lambda {\delta_B}^A + i w(A) w {\delta_B}^A
		+ {\omega_B}^A + \eps^C {C_{CB}}^A - \frac{1}{2} \xi^C {{C_C}^A}_B (-)^{BA}
\end{gather}
where we have used the notation of \cite{Butter:2009cp}.
Since the prepotentials are group elements, they must have these same
transformation properties,
and since the special conformal generator acts quite like an antiderivative,
these formulae encapsulate a good deal of information.
By inspection, one can easily see that only $V^a$ is conformally primary.
This isn't too great of a surprise, since the prepotential of conformal
supergravity is a real superfield $H^m$, and $V^a$ is its obvious quantum
variation. All other objects should in principle be given as derivatives of $V^a$
or otherwise be pure gauge artifacts.
Using the special conformal transformation rules, it is possible to rewrite each of
the prepotentials as derivatives of $V^a$ plus some remaining conformally primary
object.

As an example, note that $V^\alpha$ obeys
\[
S^\dbeta V^\alpha = -i V^{\dbeta \alpha}, \;\;\;
S_\beta V^\alpha = K_b V^\alpha = 0
\]
This is easily solved by
\[
V^\alpha = -\frac{i}{8} \nabla_\dphi V^{\dphi \alpha} + \tilde V^\alpha
\]
where $\tilde V^\alpha$ is some conformally primary superfield. The other
conditions are not all nearly so easy to solve, but the answer is straightforward
to check. One finds
\begin{align}\label{eq_V's}
V^\alpha &= -\frac{i}{8} \nabla_\dphi V^{\alpha \dphi} + \tilde V^\alpha \\
V_\dalpha &= -\frac{i}{8} \nabla^\phi V_{\phi \dalpha} + \tilde V^\dalpha \\
V(D) &=  \frac{1}{2} \nabla_c V^c + \frac{1}{2} \nabla^\alpha V_\alpha
	+ \frac{1}{2} \nabla_\dalpha V^\dalpha + \tilde V(D) \eol
	&= \frac{1}{4} \nabla_c V^c + \frac{1}{2} \nabla^\alpha \tilde V_\alpha
	+ \frac{1}{2} \nabla_\dalpha \tilde V^\dalpha + \tilde V(D) \\
V(A) &= -\frac{1}{4} \Delta_c V^c - 
	\frac{3i}{4} (\nabla^\alpha V_\alpha - \nabla_\dalpha V^\dalpha) + \tilde V(A) \eol
	&= +\frac{1}{8} \Delta_c V^c - 
	\frac{3i}{4} (\nabla^\alpha \tilde V_\alpha - \nabla_\dalpha \tilde V^\dalpha) + \tilde V(A) \\
V(M)_{\beta \alpha} &=
	+ \frac{1}{2} \nabla_{\{\beta} V_{\alpha\}}
	+ \frac{i}{8} \nabla^\dphi \nabla_{\{\beta} V_{\alpha \} \dphi} + \tilde V(M)_{\beta \alpha} \eol
	&= + \frac{1}{2} \nabla_{\{\beta} \tilde V_{\alpha\}}
	+ \frac{i}{16} \nabla^\dphi \nabla_{\{\beta} V_{\alpha \} \dphi} 
	- \frac{1}{8} \nabla_{\{\beta \dphi} {V^\dphi}_{\alpha \}} + \tilde V(M)_{\beta \alpha} \\
V(M)_{\dbeta \dalpha} &=
	+ \frac{1}{2} \nabla_{\{\dbeta} V_{\dalpha\}}
	- \frac{i}{8} \nabla^\phi \nabla_{\{\dbeta} V_{\dalpha \} \phi} + \tilde V(M)_{\dbeta \dalpha} \eol
	&= + \frac{1}{2} \nabla_{\{\dbeta} \tilde V_{\dalpha\}}
	- \frac{i}{16} \nabla^\phi \nabla_{\{\dbeta} V_{\dalpha \} \phi} 
	+ \frac{1}{8} \nabla_{\{\dbeta \phi} {V^\phi}_{\dalpha \}} + \tilde V(M)_{\dbeta \dalpha}
\end{align}
where we have defined
\begin{align}
[\nabla_\alpha, \nabla_\dalpha] \equiv -2 \Delta_{\alpha \dalpha}
\end{align}
These prepotential formulae will be the most useful to us. We have given them both in
terms of the conformally non-primary $V^\alpha$ and the primary $\tilde V^\alpha$.
The other tilded objects are similarly primary.

For completeness, we include also the special conformal prepotentials, which are
a little messier and which we will not have a great deal of use for in what follows:
\begin{align}
V(K)_\alpha &=
	+\frac{1}{8} \nabla^2 V_\alpha - \frac{1}{4} \nabla^\dphi \nabla_\alpha V_\dphi
	+ \frac{i}{96} \nabla^2 \nabla_\dphi {V_\alpha}^\dphi
	+ \frac{1}{24} \nabla_\alpha \nabla_{\beta \dbeta} V^{\beta \dbeta} + \tilde V(K)_\alpha \eol
	&= +\frac{1}{8} \nabla^2 \tilde V_\alpha - \frac{1}{4} \nabla^\dphi \nabla_\alpha \tilde V_\dphi
	+ \frac{i}{96} \nabla_\dphi \nabla^2 {V_\alpha}^\dphi
	+ \frac{1}{48} \nabla_{\{\beta} \nabla_{\alpha\} \dbeta} V^{\beta \dbeta} + \tilde V(K)_\alpha \\
V(K)_\dalpha &=
	+\frac{1}{8} \bar \nabla^2 V_\dalpha - \frac{1}{4} \nabla_\phi \nabla_\dalpha V^\phi
	+ \frac{i}{96} \bar \nabla^2 \nabla^\phi V_{\phi \dalpha}
	+ \frac{1}{24} \nabla_\dalpha \nabla_{\beta \dbeta} V^{\beta \dbeta} + \tilde V(K)_\dalpha \eol
	&= +\frac{1}{8} \bar \nabla^2 \tilde V_\dalpha - \frac{1}{4} \nabla_\phi \nabla_\dalpha \tilde V^\phi
	+ \frac{i}{96} \nabla^\phi \bar \nabla^2 V_{\phi \dalpha}
	+ \frac{1}{48} \nabla_{\{\dbeta} \nabla_{\dalpha\}\beta } V^{\beta \dbeta} + \tilde V(K)_\dalpha
\end{align}
The objects $\tilde V(K)_\alpha$ are not themselves fully primary, but are related to
$\tilde V(D)$, $\tilde V(A)$, and $\tilde V(M)_{\beta \alpha}$ by the action of
$S_\beta$. When these latter objects vanish, $\tilde V(K)_\alpha$ is itself primary.

In addition, when we consider Yang-Mills theories, we will also need
the prepotential $\Sigma^r$, the Yang-Mills prepotential associated with
the Yang-Mills generator $X_r$. It is naturally conformally primary.

We emphasize that the separation we have made above is entirely
dictated by conformality concerns; the tilded objects we have introduced
are \emph{defined} by the above equations. We will very quickly find that they are
constrained to be pure gauge artifacts. To demonstrate this, we
require two new pieces of information: the form of the chiral
gauge transformations and the first-order solution of the supergravity
constraints.

\subsection{Chiral gauge transformations}
In choosing to work in quantum Hermitian gauge, we have exhausted
the full supergroup gauge transformation, but the chiral transformations
remain. Recall they are given by
\begin{align}
U_Q \rightarrow \bar C_Q U_Q C_Q^{-1}
\end{align}
where $C_Q$ obeys a chirality condition, $[\nabla^\dalpha, C_Q]=0$.
If we define $U_Q \equiv \exp(-2i V)$, $C_Q^{-1} \equiv \exp(-2i \Lambda)$, and
$\bar C_Q \equiv \exp(-2i \bar\Lambda)$, then the above transformation rule is
equivalent (for infinitesimal $\Lambda$) to
\begin{align}
\delta V = \Lambda + \bar \Lambda - i [V,\Lambda - \bar\Lambda]
	+ \mathcal O(V^2)
\end{align}

Writing $\Lambda = \xi^A \nabla_A + \frac{1}{2} \omega^{ba} M_{ab} + \Lambda D + w A +\eps^B K_B$,
we can solve for the conditions that these various parameters must obey:
\begin{gather}
\xi_{\alpha \dalpha} = -\nabla_\dalpha L_\alpha, \;\;\;
\xi_\alpha = \frac{i}{8} \bar\nabla^2 L_\alpha, \;\;\;
\xi_\dalpha =\mathrm{arbitrary} \eol
\Lambda = -\frac{1}{2} \nabla^\dalpha \xi_\dalpha + \phi(D), \;\;\;
w = -\frac{3i}{4} \nabla^\dalpha \xi_\dalpha + \frac{i}{2} \phi(D) \eol
\omega_{\dalpha \dbeta} = \frac{1}{2} \nabla_{\{\dalpha} \xi_{\dbeta\}}, \;\;\;
\omega_{\alpha \beta} = -2i L^\gamma W_{\gamma \alpha \beta} + \phi(M)_{\alpha \beta} \eol
\eps_\dalpha = \frac{1}{8} \bar \nabla^2 \xi_\dalpha, \;\;\;
\eps_\alpha = +\frac{i}{2} L^\phi \nabla^\gamma W_{\gamma \phi \alpha} + \psi(K)_\alpha, \;\;\;
\eps_{(\alpha \dalpha)} = +i L^\phi {\nabla_\dalpha}^\gamma W_{\gamma \phi \alpha} + i \nabla_\dalpha \psi(K)_\alpha
\end{gather}
In the above formulae $\{\dalpha \dbeta\}$ denotes the (unnormalized)
symmetric sum $\dalpha \dbeta + \dbeta \dalpha$.
The superfields $\phi(D)$ and $\phi(M)_{\alpha \beta}$ are chiral,
$\psi(K)_\alpha$ is complex linear, $\xi_\dalpha$ is arbitrary, but none
of these four is primary. $L_\alpha$ is both primary and arbitrary.
As with the prepotentials, we may rewrite the non-primary
operators as derivatives of primary ones plus some new primary object.
Doing so gives
\begin{gather}
\xi_{\alpha \dalpha} = -\nabla_\dalpha L_\alpha, \;\;\;
\xi_\alpha = \frac{i}{8} \bar\nabla^2 L_\alpha, \;\;\;
\xi_\dalpha = -\frac{i}{8} \nabla_\beta \nabla_\dalpha L^\beta + \tilde \xi_\dalpha \eol
\Lambda = -\frac{1}{2} \nabla^\dalpha \xi_\dalpha - \frac{i}{16} \bar\nabla^2 \nabla_\beta L^\beta + \tilde \phi(D), \;\;\;
w = -\frac{3i}{4} \nabla^\dalpha \xi_\dalpha + \frac{1}{32} \bar\nabla^2 \nabla_\beta L^\beta
	+ \frac{i}{2} \tilde \phi(D) \eol
\omega_{\dalpha \dbeta} = \frac{1}{2} \nabla_{\{\dalpha} \xi_{\dbeta\}}, \;\;\;
\omega_{\alpha \beta} = -2i L^\gamma W_{\gamma \alpha \beta} 
	- \frac{i}{16} \bar\nabla^2 \nabla_{\{\alpha} L_{\beta\}} + \tilde \phi(M)_{\alpha \beta}
\end{gather}
We have not included the terms corresponding to $\eps(K)$ since they are fairly messy
and we don't actually have much use for these specific formulae in what follows.

The useful part of the above formulae is to note the correspondence between the
tilded gauge objects and the tilded prepotentials. For example, if we could
show that $\tilde V(K)_\alpha$ were constrained to be complex linear, then it is
a pure gauge artifact, cancelling against $\psi(K)_\alpha$. Similarly, if we
could show that $\tilde V(M)_{\alpha \beta}$ were chiral, we could cancel
it against $\tilde\phi(M)_{\alpha \beta}$. Clearly $\tilde V_\dalpha$ already
corresponds to $\tilde \xi_\dalpha$. To eliminate $\tilde V(D)$ and
$\tilde V(A)$, we would need to show that they can be related to the appropriate
sum (or difference) of a chiral and an antichiral field -- in this case,
$\tilde \phi(D)$ and its conjugate. Provided these constraints can be
enforced, the theory becomes one entirely of $V^a$.

We should check that the number of degrees of freedom work out.
$V^a$ itself consists of 32 bosonic and 32 fermionic degrees of freedom.
The gauge degree of freedom $L_\alpha$, however, \emph{also} seems to have 32+32 components.
The solution to this puzzle is that $L_\alpha$ has weight $(-3/2, -1)$ which has precisely the
ratio  necessary to accomodate a primary chiral superfield. We will find in physical models,
in fact, that $L_\alpha$ itself possesses a gauge symmetry of
$L_\alpha \rightarrow L_\alpha + \phi_\alpha$, where $\phi_\alpha$ has
8+8 components. Since it is a second order gauge degree of freedom
(ie. a gauge degree of freedom for a gauge degree of freedom), these components
contribute positively to the counting. Put more simply,
\[
32 + 32 - \left(32 + 32 - (8 + 8) \right) = 8 + 8
\]
which is the right number for conformal supergravity.
It is interesting that the physical degrees of freedom of conformal
supergravity coincide with those of a chiral spinor.

For completeness, we also include the Yang-Mills variation:
\begin{align}
\Lambda^r = i L^\beta W_\beta{}^r + \tilde \Lambda^r
\end{align}
where $\tilde \Lambda^r$ is chiral. Note that because we have included
$\Sigma^r$ with the supergravity prepotentials, its chiral gauge variation includes
a term coming from supergravity, in addition to the usual chiral superfield.

\subsection{First-order constraint solution}
We next turn to the task of solving the supergravity constraints to first order.
Because conformal supergravity is characterized by conventional constraints
as in super Yang-Mills, the curvatures are entirely described by
``gaugino'' superfields $\mathcal W_{\alpha}$ which are given by the
commutators
\begin{align}
[\nabla_\alpha, \nabla_{\beta \dbeta}] = -2i \eps_{\alpha \beta} \mathcal W_\dbeta, \;\;\;
[\nabla_\dalpha, \nabla_{\beta \dbeta}] = -2i \eps_{\dalpha \dbeta} \mathcal W_\beta
\end{align}
These are superfields which obey a chirality condition,
$\{\nabla_\dalpha,\mathcal W_\beta\} = 0$. The constraints of conformal
supergravity involve imposing
$\mathcal W_{\alpha}(P)^B=\mathcal W_{\alpha}(D)=\mathcal W_{\alpha}(A)=0$.
From these it follows that $\mathcal W_{\alpha}(M)^{\dbeta \dgamma}=0$ and
$\mathcal W_\alpha(K)_\dalpha=0$ and that all the remaining curvatures can
be expressed in terms of the single chiral superfield $W_{\alpha \beta \gamma}$.

The chiral superfield $\mathcal W_\alpha$ can be defined by
\begin{align}
8 \mathcal W_\alpha = [\nabla_\dalpha, \{\nabla^\dalpha, \nabla_\alpha\}]
	= +2i  [\nabla^\dalpha, \nabla_{\alpha \dalpha}]
\end{align}
Varying this object to first order involves varying each of the covariant
derivatives on the right side. The easiest way to handle this is to adopt
a chiral quantum gauge where we force all of the quantum variation onto
$\nabla_\alpha$ and leave $\nabla_\dalpha$ unchanged. If the gaugino superfield
vanishes in this gauge, it vanishes in any gauge, including quantum Hermitian gauge.
(This is equivalent to doing the variation in Hermitian gauge and then performing a quantum
prepotential-dependent gauge transformation.)

Thus,
\begin{gather}
\delta_c \nabla_\alpha = [2i V, \nabla_\alpha], \;\;\; \delta_c \nabla_\dalpha = 0
\end{gather}
where the subscript $c$ denotes that the quantum gauge is chiral.

Note first that the Hermitian quantum variation of $\nabla_\alpha$ is
\begin{align}
\delta \nabla_\alpha = [i V, \nabla_\alpha] \equiv - H_\alpha{}^B X_B
	= - {H_\alpha}^B \nabla_B - \Omega_\alpha(M) - \Lambda_\alpha D
	- \omega_\alpha A - J_\alpha{}^B K_B
\end{align}
where
\begin{align}
{H_\alpha}^\beta &= +i \nabla_\alpha V^\beta - i {V(M)_\alpha}^\beta
	- \frac{i}{2} V(D) {\delta_\beta}^\alpha
	- V(A) {\delta_\beta}^\alpha \\
H_{\alpha \dbeta} &= +i \nabla_\alpha V_\dbeta \\
H_{\alpha (\beta \dbeta)} &= +i \nabla_\alpha V_{(\beta \dbeta)} + 4 \eps_{\alpha \beta} V_{\dbeta} \\
\Omega_\alpha(M) &= +i V^b R_{b \alpha}(M) + i \nabla_\alpha V(M) + 2i V(K)^\beta M_{\beta \alpha} \\
\Lambda_\alpha &= +i \nabla_\alpha V(D) + 2i V(K)_\alpha \\
\omega_\alpha &= +i \nabla_\alpha V(A) + 3 V(K)_\alpha \\
{J_\alpha}^\beta &= +i \nabla_\alpha V(K)^\beta \\
J_{\alpha \dbeta} &= +i \nabla_\alpha V(K)_\dbeta + i V^c R_{c \alpha}(K)_\dbeta + V(K)_{\alpha \dbeta} \\
{J_{\alpha}}^b &= +i \nabla_\alpha V(K)^b + i V^c R_{c \alpha}(K)^b
\end{align}

In the chiral gauge we are using, the variation of $\nabla_\alpha$ is simply twice this:
\begin{align}
\delta_c \nabla_\alpha = -2 {H_\alpha}^B \nabla_B - 2{H_\alpha}^{\ul b} X_{\ul b}
\end{align}
The variation of the bosonic derivative is rather easy to calculate in
chiral gauge. One finds
\begin{align}
\delta_c \nabla_{\alpha \dalpha} = -i \nabla_\dalpha {H_\alpha}^B X_B
	- i \nabla_\dalpha {H_\alpha}^{\ul b} X_{\ul b}
	- 2 {H_\alpha}^\beta \nabla_{\beta \dalpha}
	+ H_{\alpha (\beta \dalpha)} \mathcal W^\beta
	+ i {H_\alpha}^{\ul b} {f_{\ul b \dalpha}}^D X_D
\end{align}
$\delta \mathcal W$ is then given by
\begin{align}
4 \, \delta \mathcal W_\alpha =&
	- \bar\nabla^2 {H_\alpha}^B X_B
	+ 4i \nabla_\dalpha {H_\alpha}^\beta {\nabla_\beta}^\dalpha \eol
	& + \left(2i \nabla^\dbeta H_{\alpha (\beta \dbeta)} + 8 H_{\alpha \beta} \right) \mathcal W^\beta \eol
	&+ \left(
		2 \nabla_\dalpha {H_\alpha}^{\ul b} - {H_\alpha}^{\ul c} {f_{\ul c \dalpha}}^{\ul b}
	\right) {f_{\ul b}}^{\dalpha D} X_D
\end{align}

We begin the analysis by considering the constraints imposed on the prepotentials by
$\mathcal W_\alpha(P)=0$. These amount to two conditions, which we write as
\begin{gather}\label{eq_cond1}
\bar \nabla^2 H_{\alpha (\beta \dbeta)} = 8i \nabla_\dbeta H_{\alpha \beta} \\ \label{eq_cond2}
8 J_{\alpha \dalpha} = -\bar\nabla^2 H_{\alpha \dalpha} - \nabla_\dalpha \Lambda_\alpha
	- 2i \nabla_\dalpha \omega_\alpha
	+ 2 \nabla^\dbeta \Omega_\alpha{}_\dbeta{}_\dalpha
\end{gather}
The second of these amounts to a definition of $V(K)_{\alpha \dalpha}$,
on which $J_{\alpha \dalpha}$ linearly depends. (There is a third condition that we haven't
listed which is a trivial consequence of the first.)

Choosing $\mathcal W_\alpha(D)$ and $\mathcal W_\alpha(A)$ to vanish amount to
the condition
\begin{gather}\label{eq_cond3}
\bar\nabla^2 \Lambda_\alpha = -\frac{2i}{3} \bar\nabla^2 \omega_\alpha
\end{gather}

All other conditions on the $\mathcal W_\alpha$'s follow from these three.

The third condition, \eqref{eq_cond3}, is the easiest to immediately evaluate.
Using the above definitions for $\Lambda_\alpha$ and $\omega_\alpha$ leads to
\[
0 = \bar\nabla^2 \left(i \nabla_\alpha V(D) - \frac{2}{3} \nabla_\alpha V(A) + 4i V(K)_\alpha \right)
\]
Inserting the definitions of the $V$'s in terms of the $\tilde V's$, we discover a nice
surprise. The above condition reduces to
\begin{align}\label{eq_VK1}
0 = \bar\nabla^2 \left(i \nabla_\alpha \tilde V(D) - \frac{2}{3} \nabla_\alpha \tilde V(A)
	+ 4i \tilde V(K)_\alpha \right)
\end{align}

The first condition, \eqref{eq_cond1}, is the next easiest to
check. Again using the $\tilde V$'s we can conclude that
\begin{gather}
0 = \nabla_\dbeta \tilde V(M)_{\beta \alpha} \\
0 = \nabla_\dbeta \left(\frac{i}{2} \tilde V(D) + \tilde V(A) \right)
\end{gather}
The first of these implies that $\tilde V(M)_{\beta \alpha}$ is chiral and
therefore pure gauge: it is in one-to-one correspondence with its chiral
gauge parameter $\tilde \phi(M)_{\beta \alpha}$. We can therefore choose
$\tilde V(M)$ to vanish. The second equation implies that
\[
\tilde V(D) - 2i \tilde V(A) = 2 \tilde \phi(D)
\]
Together with its conjugate, this implies that $\tilde V(D)$ and $\tilde V(A)$
are the real and imaginary parts of a chiral superfield $\tilde \phi(D)$. Since this
also precisely overlaps with their gauge degrees of freedom, we can similarly
choose $\tilde V(D)$ and $\tilde V(A)$ to vanish.

This last point is an important one. In a theory with a conformal
compensator $\Phi_0$ of unit scaling dimension and matter fields
$\Phi^i$ of vanishing scaling dimension, the quanta of $\Phi_0$
are indistinguishable from the chiral degree of freedom $\tilde\phi(D)$.
Both have an equally valid claim to be the chiral quanta which
together with $V^a$ make up the quanta of Poincar\'e
supergravity, while the other is the pure gauge degree of freedom.
From our point of view, it is almost always more sensible to remove
$\tilde \phi(D)$ immediately. If desired, it can be restored by undoing
the chiral scale transformation.

Whether or not we choose to eliminate $\tilde\phi(D)$, the condition that
$\tilde V(D)$ and $\tilde V(A)$ are made up of a sum and a difference
of a chiral and an antichiral superfield together with \eqref{eq_VK1}
implies that
\begin{align}
\bar\nabla^2 \tilde V(K)_\alpha = 0
\end{align}
This means that $\tilde V(K)_\alpha$ is a complex linear superfield
and so it too is in perfect correspondence with its gauge degree of
freedom and so can be taken to vanish.

We return now to the second condition, \eqref{eq_cond2}. This boils down
to
\begin{align}
V(K)_{\alpha \dalpha} = -i \nabla_\alpha V(K)_\dalpha - i \nabla_\dalpha V(K)_\alpha
	+ \frac{i}{8} \nabla_\alpha \bar\nabla^2 V_\dalpha
	+ \frac{i}{8} \nabla_\dalpha \bar\nabla^2 V_\alpha
	+ \frac{1}{32} \hat \Delta_D V_{\alpha \dalpha}
\end{align}
where we have defined
\begin{align}
\hat \Delta_D V_{\alpha \dalpha} =
	\nabla^\beta \bar\nabla^2 \nabla_\beta V_{\alpha \dalpha}
	+ 16 \nabla_\dgamma W^{\dgamma \dbeta}{}_{\alpha} V_{\alpha \dbeta}
	+ 16 W_\alpha{}^{\beta \gamma} \nabla_\gamma V_{\beta \dalpha}
\end{align}
One can show that $\hat \Delta_D V_a$ is Hermitian.

Before moving on, we note here the chiral variation of the conformal supergravity
field strength in the chiral gauge where $\tilde V(D)$, $\tilde V(M)$, and $\tilde V(A)$
vanish:
\begin{align}
\delta_c W_{\alpha \beta \gamma} = \sum_{(\alpha \beta \gamma)}
	\frac{i}{96} \bar\nabla^2 \nabla^\dphi{}_\alpha \nabla_\beta V_{\gamma \dphi}
\end{align}

We have discovered how to use the Yang-Mills-like features of the conformal
supergravity algebra to extract the geometric quanta at first order. We turn
next to some specific physical models.

\section{Two physical models at first order}
\subsection{Linear compensator model}
Although we will be most concerned with an arbitrary chiral
model, we will first consider a simpler model.
The minimally coupled linear compensator model with a K\"ahler potential
consists of a D-term action of two terms
\begin{align}
S = S_{G} + S_K.
\end{align}
The Einstein-Hilbert term is contained within the first term
\begin{align}
S_G = \int E L V_R \equiv 3 \int E L \log (L / \Phi_0 \bPhi_0)
\end{align}
where $L$ is the linear compensator and $\Phi_0$ is a chiral superfield of
scaling dimension 1, whose presence is almost solely to make the
argument of the logarithm conformally invariant, as a redefinition
\[
\Phi_0 \rightarrow e^{\Lambda} \Phi_0
\]
for chiral $\Lambda$ leaves the action invariant due to the linearity
condition of $L$. In the gauge where $L=1$, this has the form of a
Fayet-Iliopoulos for the supergravity $U(1)_R$.

The coupling of chiral matter to the theory is contained within the second term
\begin{align}
S_K = \int E L K
\end{align}
where $K$ is the K\"ahler potential, a dimension zero Hermitian function of chiral
and antichiral superfields which possesses a symmetry
\begin{align}
K \rightarrow K + F + \bar F,
\end{align}
also a consequence of the linearity of $L$.

We could also include Fayet-Iliopoulos terms for Yang-Mills fields by introducing them as
$\int E L \Tr V$ where $V$ is the gauge prepotential. In fact, one can likewise view $S_K$
as essentially being the FI term for a $U(1)_K$ symmetry. One would then naturally
combine all these to give the single term
\begin{align}
-3 \int E \,L \log \left(\Phi_0 e^{-(K+V)/3} \bPhi_0 / L\right)
\end{align}
which can be understood as a sum of the FI terms for the Yang-Mills, K\"ahler, and
$U(1)_R$ gauge sectors. We will exclude from our discussion Yang-Mills FI terms
and treat the supergravity and K\"ahler sectors separately.

In order to proceed, we need to determine the transformation of the various quantities.
We will work in the gauge where $\tilde V(D) = \tilde V(A) = \tilde V(M) = \tilde V(K) = 0$.
The non-primary object $V^\alpha$ we will leave for the moment unfixed and specify a gauge
for it later.

The first order variation of $E$ is
\begin{align}
\delta E &= {H^\alpha}_\alpha + {H_\dalpha}^\dalpha + {H^a}_a \eol
	&= -3i \nabla^\alpha V_\alpha + 3i \bar\nabla_\dalpha V^\dalpha - \Delta_b V^b
	- 4 V(A) = 0
\end{align}
This is an initially surprising
result, but it is owed to our working in a conformal theory. For example, in a component
four dimensional theory, the first order variation of $\sqrt{g}$ is the trace of the
graviton perturbation, which is the conformal mode of the graviton. We could set the
scaling gauge in such a theory by forcing the conformal mode to vanish. This is something
of a shell game, however, since the conformal mode of the graviton is essentially the
same object as the conformal compensator in such a theory. In the current theory, the
role of the ``conformal mode'' of the graviton will be taken up by the linear compensator
(and later the chiral compensator) and so $\delta E  = 0$ here.

The first order variation of a chiral superfield $\Phi$ of scaling dimension $\Delta$ and
$U(1)_R$ weight $2\Delta / 3$ is given in Hermitian gauge by
\begin{align}
\delta \Phi &= -i V^B X_B \Phi + \delta_c \Phi \eol
	&= -i V^\beta \Phi - i V^b \nabla_b \Phi - i \left(V(D) + \frac{2i}{3} V(A)\right) \Delta \Phi
	-i \Sigma^r X_r \Phi + \eta
\end{align}
where we define $\delta_c \Phi \equiv \eta$ as the variation in chiral gauge.

We next note that $L$ may be written
\begin{align}
L = \nabla^\alpha \Phi_\alpha + \nabla_\dalpha \Phi^\dalpha
\end{align}
in terms of chiral primary superfields $\Phi_\alpha$ of weight $(3/2,1)$.
The variation of $\nabla^\alpha \Phi_\alpha$ is given by
\begin{align}
\delta (\nabla^\alpha \Phi_\alpha) =
	& -i \nabla^\beta (V_\beta \nabla^\alpha \Phi_\alpha)
	+i \nabla_\dbeta (V^\dbeta \nabla^\alpha \Phi_\alpha)
	- \Delta_b (V^b \nabla^\alpha \Phi_\alpha)
	+ 2 V^{\dalpha \alpha} \bar W_\dalpha \Phi_\alpha \eol
	& + \frac{1}{4} \nabla_\dalpha \nabla^2 (V^{\dalpha \alpha} \Phi_\alpha)
	+ \nabla^\alpha (\delta_c \Phi_\alpha) \eol
	& - i \Sigma \nabla^\alpha \Phi_\alpha - 2i (\nabla^\alpha \Sigma^r) X_r \Phi_\alpha
\end{align}
Assuming $\Phi_\alpha$ to be a gauge singlet, we can write the variation
of $L$ as
\begin{align}
\delta L = \mathcal L - i \nabla^\beta (V_\beta L) + i \nabla_\dbeta (V^\dbeta L) - \Delta_b (V^b L)
\end{align}
where
\begin{align}
\mathcal L \equiv \nabla^\alpha \left(\delta_c \Phi_\alpha -
	\frac{1}{4} \bar\nabla^2 (V^{\dalpha \alpha} \bPhi_\dalpha)\right) + \hc
	\equiv \nabla^\alpha \eta_\alpha + \hc
\end{align}
$\eta_\alpha$ is a weight $(3/2,1)$ chiral primary superfield, which we have defined
to depend on both $\Phi_\alpha$ and $\bPhi^\dalpha$ so as to simplify the formula.

After several integrations by parts, one can show that
\begin{align}
\delta S_G = \int E \left( \mathcal L V_R - 2 V^b \Delta_b L + \frac{3}{2L} V^{\alpha \dalpha} \nabla_\alpha L \nabla_\dalpha L\right)
\end{align}
We may define a new weight (0,0) primary superfield $G_b$ by
\begin{align}
G_b \equiv \frac{1}{2} L^{-1} \Delta_b L - \frac{3}{8L^2} \nabla_\alpha L \nabla_\dalpha L
	= - L^{1/2} \Delta_b L^{-1/2}
\end{align}
So that
\begin{align}
\delta S_G = \int E \left( \mathcal L V_R - 4 L V^b G_b \right)
\end{align}

One can similarly work out the structure of $S_K$. Skipping details
(the most difficult of which is an integration by parts) one finds
\begin{align}
\delta S_K = \int E L \left( K_i \eta^i + K_{\bj} \eta^\bj + V^b K_b + \Sigma^r K_r\right) + \int E \mathcal L K
\end{align}
where
\begin{gather}
K_{\alpha \dalpha} \equiv K_{i \bj} \nabla_\alpha \Phi^i \nabla_\dalpha \bPhi^\bj \\
K_r \equiv -i K_i X_r \Phi^i + i K_\bj X_r \bPhi^\bj
\end{gather}
Both $K_a$ and $K_r$ are conformally primary.

Combining these two variations gives
\begin{align}
\delta S = \int E \, \left[ L V^b \left(-4 G_b + K_b \right)
	+ L \Sigma^r K_r
	+ L K_i \eta^i + L K_{\bj} \eta^\bj
	+ \mathcal L (V_R + K) \right]
\end{align}
This is a surprisingly compact expression. When $L$ is gauged to 1, $G_b$ becomes
the Poincar\'e superfield of the same name and represents the pure supergravity
contribution to the energy-momentum tensor. $K_b$ represents the matter contribution
to the energy-momentum tensor, and $K_r$ is the matter contribution to the gauge current.

\subsubsection{Gauge invariance of the linear compensator model}
The first feature we should observe about our linear compensator model is that
at first order it is independent of $V^\alpha$ and $V_\dalpha$. This is certainly
sensible since these are gauge degrees of freedom and should certainly not
have any equations of motion associated with themselves.

The dynamical theory would seem to consist of $V^a$ and $\Sigma^r$ -- the
Hermitian superfields associated with the graviton and gauge multiplets --
as well as the matter superfield $\eta^i$ and $\bar\eta^\bj$ and the
linear compensator variation $\mathcal L$. We recall that $V^a$ transforms
under the quantum chiral gauge transformation as
\begin{align}
\delta V_{\alpha \dalpha} = \nabla_\alpha L_\dalpha - \nabla_\dalpha L_\alpha
\end{align}
Under the $L_\alpha$ transformation, a chiral superfield transforms as
\begin{align}
\Phi' = C_Q \Phi
\end{align}
Differentially, this reads
\begin{align}
\delta \eta = 2i \Lambda \Phi =
	2i \xi^a \nabla_a \Phi
	+ 2i \xi^\alpha \nabla_\alpha \Phi
	+ 2i \Lambda \Delta \Phi
	-\frac{4}{3} \omega \Delta \Phi
	+ 2i \Lambda^r X_r \Phi
\end{align}
where $\Delta$ is the scaling dimension of $\Phi$.
Plugging in the values for superfields, we find
\begin{align}
\delta \eta = -\frac{1}{4} \bar\nabla^2 \left(L^\alpha \nabla_\alpha \Phi\right)
	- \frac{\Delta}{12} (\bar\nabla^2 \nabla^\beta L_\beta) \Phi
	+ 2i \tilde \Lambda^r X_r \Phi
\end{align}
The gauge superfield $\Sigma^r$ transforms as
\begin{align}
\delta \Sigma^r = \tilde \Lambda^r + \bar{\tilde\Lambda}^r
	+ i L^\beta W_\beta^r + i L_\dbeta \bar W^{\dbeta r}
\end{align}
The quantum linear compensator varies as
\begin{align}
\delta \mathcal L = \frac{1}{4} \nabla^\alpha \bar\nabla^2 (L_\alpha L) + \hc
\end{align}
Note that this last expression depends on $\Phi_\alpha$ only implicitly via $L$.

One can check that the first-order action is invariant under this first-order
shift in the quantum superfields, as it must be by construction.

\subsection{Arbitrary chiral model}
The minimal linear compensator model is notable for the clean decoupling of the
gravitational and matter terms of the action, which gives a corresonding decoupling
of their contributions to the gravitational current. The arbitrary chiral model
will not be so immediately simple to evaluate, but we will find its first order
variation shares the same features.

The chiral model classically dual to the minimal linear compensator model 
with a K\"ahler potential $K$ is
\begin{align}
S = -3 \int E \,\Phi_0 \bPhi_0 \, e^{-K/3}
\end{align}
This action encapsulates not only the pure gravity effects (denoted
$S_G$ in the linear model) but also kinetic matter terms (denoted $S_K$).
Here $\Phi_0$ is a weight $(1,2/3)$ conformally primary chiral superfield
and $K$ is as before a Hermitian function of weight $(0,0)$ chiral and
antichiral superfields. A canonically normalized Einstein-Hilbert
term is found in the gauge $\Phi_0 \bPhi_0 = e^{K/3}$.

The above D-term is a special case of a more general theory involving
an arbitrary set of chiral superfields of arbitrary weights,
\begin{align}
S = -3\int E Z \equiv -3 \left[ Z \right]_D
\end{align}
We have introduced the shorthand that $[ \;\;\;]_D$ denotes integration
of its argument over the full superspace. We can similarly define
$[ \;\;\; ]_F$ as integration over the chiral submanifold of superspace.
In this expression, $Z$ is
a gauge invariant Hermitian superfield of scale dimension two construced
from the chiral superfields $\Phi^i$ and their conjugates. The factor
of -3 is necessary so that the gauge $Z=1$ gives a canonical Einstein-Hilbert
term. The proof of this is straightforward. Using the scaling and
$U(1)_R$ weights of $Z$,
\begin{gather*}
D Z = 2 Z = Z_i \Delta_i \Phi^i
	+ Z_{\bj} \Delta_{\bj} \bPhi^\bj \\
-\frac{3i}{2} A Z = 0 = Z_i \Delta_i \Phi^i
	- Z_{\bj} \Delta_{\bj} \bPhi^\bj 
\end{gather*}
and that the Einstein-Hilbert term is contained within
\[
-3 [Z]_D = -3\left[Z_\bj \CP \Phi^\bj + \ldots\right]_F
	= -3 Z_{\bj} \ACP \CP \Phi^\bj + \ldots
	= -3 Z_\bj \Box \Phi^\bj + \ldots
\]
where $\CP = -\bar\nabla^2 / 4$, $\ACP = -\nabla^2/4$ and $\Box$ are
superconformal. That $\Box$ is superconformal means it contains
$\mathcal R/6$ weighted by the scaling dimension of the field on
which it acts, and so it is easy to see that the Einstein-Hilbert term is
\[
-3 [Z]_D \ni -\frac{1}{2} \mathcal R Z_\bj \Delta_\bj \Phi^\bj
	= -\frac{Z}{2} \mathcal R
\]
The gauge $Z=1$ then corresponds to a canonical Einstein-Hilbert term.

Since $\delta E = 0$, we concern ourselves only with the first order variation of $Z$:
\begin{align}
\delta Z = & Z_i (\eta^i - i V \Phi^i) + Z_\bj (\bar\eta^\bj + i V \bPhi^\bj) \eol
	= & Z_i \eta^i + Z_\bj \bar\eta^\bj
		- i Z_i \Sigma^r X_r \Phi^i + i Z_\bj \Sigma^r X_r \bPhi^\bj
		- i Z_i V^b \nabla_b \Phi^i + i Z_\bj  V^b \nabla_b \bPhi^\bj \eol
		& - i V^\alpha \nabla_\alpha Z + i V_\dalpha \bar \nabla^\dalpha Z
		+ \frac{4}{3} V(A) Z
\end{align}
Plugging in the value of $V(A)$ gives
\begin{align}
\delta Z = & Z_i \eta^i + Z_\bj \bar\eta^\bj
		- i Z_i \Sigma^r X_r \Phi^i + i Z_\bj \Sigma^r X_r \bPhi^\bj
		- i Z_i V^b \nabla_b \Phi^i + i Z_\bj  V^b \nabla_b \bPhi^\bj \eol
		& + i \nabla_\alpha (V^\alpha Z) - i \bar \nabla^\dalpha (V_\dalpha Z)
		- \frac{1}{3} \Delta_b V^b Z
\end{align}

The two terms in the last line which appear to vanish as total derivatives actually
do not. To see why, note that
the actual statement of a vanishing total derivative involves only the coordinate
derivative:
\[
0 = \partial_M ( E {E_\alpha}^M V^\alpha Z ) =
	\nabla_M ( E {E_\alpha}^M V^\alpha Z ) + {h_M}^{\ul b} X_{\ul b} (E {E_\alpha}^M V^\alpha Z)
\]
The term involving the connection usually vanishes by gauge invariance; however,
in this case $V^\alpha$ is not conformally invariant (though the other terms in the
parentheses are), and so the second term yields
\[
E f_{\alpha \dalpha} \bar S^\dalpha (V^\alpha Z)= E \left(-i f_{\alpha \dalpha} V^{\alpha \dalpha} Z\right)
\]
Evaluating the first term yields
\[
E \left( \nabla_\alpha (V^\alpha Z)+ {T_{\alpha B}}^B V^\alpha Z\right)
\]
The trace of the torsion tensor vanishes, which leads to the identity
\[
i \nabla_\alpha (V^\alpha Z)= - f_{\alpha \dalpha} V^{\alpha \dalpha} Z + \textrm{t.d.}
\]
Integrating by parts on the $\Delta_b V^b$ term gives the same explicit connections
but with the opposite sign, yielding
\begin{align}
\delta S = & -3 Z_i \eta^i -3 Z_\bj \bar\eta^\bj
		+ 3 i Z_i \Sigma^r X_r \Phi^i - 3i Z_\bj \Sigma^r X_r \bPhi^\bj 
		 + V^b \left(\Delta_b Z + 3i Z_i \nabla_b \Phi^i - 3i Z_\bj \nabla_b \bPhi^\bj\right)
\end{align}
There are several annoying features of this expression.
One is that the terms involving $V^b$ are not individually conformally invariant.
Another is that in the linear compensator model, we had a clear factor of $L$ out front
of all the terms which we could gauge to one. Here we would like to gauge $Z=1$
to arrive at the supergravity of Binetruy, Girardi, and Grimm \cite{bgg},
but none of the terms possess an explicit
$Z$ out front. We can deal with both of these issues by the following
field redefinition:
\begin{align}
\K \equiv -3 \log Z
\end{align}
$\K$ is a superfield which transforms non-linearly under a conformal transformation.
If we choose $Z = \Phi_0 \bPhi_0 e^{-K/3}$, we see that this $\K$ is
essentially the same object as the canonical K\"ahler potential:
\[
\K = K - 3 \log(\Phi_0 \bPhi_0)
\]
The advantage of this definition is that we may now rewrite $\delta S$ as
\begin{align}
\delta S = Z \left(
		\K_i \eta^i + \K_\bj \bar\eta^\bj
		+ \Sigma^r \K_r
		+ V^b \left(-4 G_b + \K_b \right)
		\right)
\end{align}
where we have defined
\begin{gather}
G_b \equiv - Z^{1/2} \Delta_b Z^{-1/2} \\
\K_{\alpha \dalpha} \equiv \K_{i \bj} \nabla_\alpha \Phi^i \nabla_\dalpha \bPhi^\bj \\
\K_r \equiv -i \K_i X_r \Phi^i + i \K_\bj X_r \bPhi^\bj
\end{gather}

If we choose $Z = \Phi_0 \bPhi_0 e^{-K/3}$, then we find
\begin{align}
\delta S = Z \left(
		K_i \eta^i + K_\bj \bar\eta^\bj
		+ \Sigma^r K_r
		+ V^b \left(-4 G_b + K_b \right)
		- \frac{3 \eta_0}{\Phi_0} - \frac{3 \bar \eta_0}{\bPhi_0}
		\right)
\end{align}
and the chiral first-order action is superficially the same as the linear one
except for the exchange of the $\mathcal L$ sector for the $\eta_0$ sector
and the exchange of the $L$ compensator for $Z$.

The importance of this observation is that it simplifies the
task of finding the second-order action for both of these theories.
Rather than treating each individually, we can focus on their common
features and only worry about where they specifically differ.

Let us consider several other terms that we might like to include
in both of these models.

\subsection{Superpotential terms}
A superpotential term is a chiral action $S_P$ defined as
\begin{align}
S_P = \int \chE\, P + \hc
\end{align}
where $P$ is some chiral superfield of weight $(3,2)$. For the
simplest chiral compensator model, $P = \Phi_0^3 W$ where $W$ is the
object one normally calls the superpotential. Because we're interested
in linear compensator models as well as the general chiral model,
we will use the more generic name $P$ to denote this F-term superfield
Lagrangian.

Since the superpotential terms involve purely chiral and antichiral actions,
we can use the quantum chiral and antichiral gauges to describe
them. We note that
\begin{align}
\delta_c \chE &= {H^\alpha}_\alpha + {H^a}_a = 0
\end{align}
in quantum chiral gauge, so only the chiral variation of the integrand remains.
The variation of the superpotential term is then simply
\begin{align}
\delta_c S_P = \int \chE\, P_i \eta^i + \hc
\end{align}
implying that the superpotential plays no rule in the pure conformal supergravity
equations of motion. (That it plays a role in Poincar\'e supergravity arises
because of the presence of the chiral compensator.)

\subsection{Yang-Mills terms}
The Yang-Mills term we will consider is
\begin{align}
S_{YM} = \frac{1}{4} \int \chE\, f_{rs} W^{\alpha r} W_{\alpha}{}^s + \hc
\end{align}
where $f_{rs}$ is a holomorphic covariant gauge coupling. In the simplest
of cases, $f_{rs} = \delta_{rs}$, but we will for the moment
allow for a more generic holomorphic coupling.

As before, one finds quantum chiral gauge the simplest for the chiral action.
Using
\begin{align}
\delta_c W_\alpha^r = -\frac{i}{4} \bar\nabla^2 \nabla_\alpha \Sigma^r
	-\frac{1}{4} \bar\nabla^2 \left( V_{\alpha \dbeta} \bar W^{\dbeta r}\right)
\end{align}
as well as
\begin{align}
\delta_c f_{rs} = f_{rs,i} \eta^i
\end{align}
one immediately finds
\begin{align}
\delta S_{YM} &= \int \chE \left(\frac{1}{4} f_{rs,i} \eta^i W^{\alpha r} W_{\alpha}{}^s
	- \frac{i}{8} f_{rs} W^{\alpha r} \bar\nabla^2 \nabla_\alpha \Sigma^s
	- \frac{1}{8} f_{rs} W^{\alpha r} \bar\nabla^2 (V_{\alpha \dbeta} \bar W^{\dbeta s})
	 \right) + \hc \eol
	&= \int \chE \left(\frac{1}{4} f_{rs,i} \eta^i W^{\alpha r} W_{\alpha}{}^s\right)+
	\int E\left( \frac{i}{2} f_{rs} W^{\alpha r} \nabla_\alpha \Sigma^s
	+ \frac{1}{2} f_{rs} V_{\alpha \dalpha} W^{\alpha r} \bar W^{\dalpha s})\right) + \hc
\end{align}

There is the possibility of introducing the Yang-Mills interactions
by requiring the linear compensator $L$ to obey the modified linearity conditions
\[
\bar \nabla^2 L = 2k \Tr (W^\alpha W_\alpha), \;\;\;
\nabla^2 L = 2k \Tr(\bar W_\dalpha \bar W^\dalpha)
\]
Then Yang-Mills interactions can be made part of the structure of superspace when
the compensator is gauged to 1. This tends to introduce non-holomorphic
gauge couplings. We will avoid this possibility for now and
restrain ourselves to the normal holomorphic Yang-Mills terms.

\subsection{Generic first-order structure}
We summarize the generic structure that the arbitrary chiral model and the
minimal linear compensator models possess. The common part of the first order action
consists of a sum of four terms. They are:
\begin{align}
(\delta S)_G =& \left[-4 X V^b G_b\right]_D \\
(\delta S)_K =& \left[X (V^b \K_b + \Sigma^r \K_r + \eta^i \K_i + \bar\eta^\bj \K_\bj)\right]_D \\
\delta S_P =& \left[\eta^i P_{i} \right]_F + \hc \\
\delta S_{YM} =& \left[V^a \mathcal Y_a + \Sigma^r \mathcal Y_r \right]_D
	+ \left[\eta^i \mathcal Y_i \right]_F
	+ \left[\bar\eta^\bj \bar {\mathcal Y}_\bj\right]_{\bar F}
\end{align}
where $X$ is the compensator ($L$ or $Z$) and
\begin{gather}
G_b \equiv - X^{1/2} \Delta_b X^{-1/2} \\
\K_{\alpha \dalpha} \equiv \K_{i \bj} \nabla_\alpha \Phi^i \nabla_\dalpha \bPhi^\bj \\
\K_r \equiv -i \K_i X_r \Phi^i + i \K_{\bj} X_r \bPhi^\bj \\
\mathcal Y_i \equiv \frac{1}{4} f_{rs,i} W^{\alpha r} W_{\alpha}{}^s \\
\mathcal Y_{\alpha \dalpha} \equiv - (f_{rs} + \bar f_{rs}) W_\alpha{}^r \bar W_\dalpha{}^s \\
\mathcal Y_r \equiv -\frac{i}{2} \nabla^\alpha \left(f_{rs} W_\alpha{}^s \right) + \hc
\end{gather}
We will find use to denote $G_{rs} \equiv f_{rs} + \bar f_{rs}$. Then the last two
equations above may be written
\begin{gather*}
\mathcal Y_{\alpha \dalpha} \equiv - G_{rs} W_\alpha{}^r \bar W_\dalpha{}^s \\
\mathcal Y_r \equiv -\frac{i}{2} (\nabla^\alpha G_{rs}) W_\alpha{}^s
	- \frac{i}{2} (\nabla_\dalpha G_{rs}) \bar W^\dalpha{}^s
	- \frac{i}{2} G_{rs} \nabla^\alpha W_\alpha{}^s
\end{gather*}
using $\nabla^\alpha W_\alpha{}^r = \nabla_\dalpha \bar W^\dalpha{}^r$.

The equations of motion amount to
\begin{align}
0 &= -4 X G_b + X \K_b + \mathcal Y_b \\
0 &= \K_r + \mathcal Y_r \\
0 &= -\frac{1}{4} \bar\nabla^2 (X \K_i) + P_i + \mathcal Y_i
\end{align}

For the linear compensator model, there is the additional term
\begin{align}
\delta S_L = \left[\mathcal L (V_R + K) \right]_D
\end{align}
along with that model's equation of motion
\begin{align}
0 = \bar\nabla^2 \nabla_\alpha(V_R + K) = \nabla^2 \bar\nabla^\dalpha(V_R + K)
\end{align}
which implies that $V_R = -K$ up to the real part of a chiral superfield.

The structure we have identified here is actually more general than this treatment
indicates. The same features persist in arbitrary models involving any number of
linear and chiral superfields. A brief discussion of the first order
variation of an \emph{arbitrarily} coupled linear superfield is given in
Appendix \ref{arblin}.

\section{Going to second order}
In order to construct a one-loop effective action, we require the action
to second order in the quantum deformations. The simplest way to do this
is a sort of bootstrap: vary our first order expression again to first order.

However, doing so immediately tends to produce a nasty set of terms involving
many derivatives of the compensator $X$ for the graviton's action.
The reason is easy to see: the action for the graviton is hidden within
the action for the compensator. In addition to a term
$X V^a \Box V_a$, there would be a host of terms involving derivatives
of $X$ needed in order to make this expression invariant under
special conformal transformations. One way to simplify this would be to eliminate
many of these terms by choosing a gauge where $X$ is constant and then degauging
to Poincar\'e derivatives. Unfortunately this sacrifices
the conformal invariance of the classical action before quantization has
even taken place. A better approach would be to introduce conformally
\emph{invariant} derivatives, with respect to which $X$ is covariantly
constant. These would compactly encode the many terms involving derivatives
of $X$ in conformally invariant combinations. It is to this construction
that we now turn.

\subsection{A brief interlude: conformally invariant (or compensated) derivatives}
\subsubsection{Definition}
In the preceding discussion, we introduced the conformally primary superfield
$G_b$ which was defined in terms of the dimension 2 compensator $X$. When $X$
is gauged to unity and the conformally covariant derivatives are themselves
``degauged'', the object $-X^{1/2} \Delta_b X^{-1/2}$ reduces simply to the Poincar\'e
superfield $G_b$, but the existence of this conformally primary combination
means we may identify the equivalent of $G_b$ even in the conformal
theory. We may similarly identify other Poincar\'e equivalents and thereby
perform something very much like a degauging while still maintaining the
underlying conformal invariance.

We begin with $X$, a primary Hermitian superfield with $\Delta=2$ and $w=0$. Define
$U = \log X$ so that under scalings, $U$ transforms nonlinearly into a constant, here
$D U = 2$. Then we define the compensator-assisted derivatives as
\begin{align}
\CD_\alpha \equiv \nabla_\alpha - \frac{1}{2} \nabla_\alpha U D - \frac{1}{2} \nabla^\beta U M_{\beta \alpha}
	+ \frac{3i}{4} \nabla_\alpha U A \\
\CD^\dalpha \equiv \nabla^\dalpha - \frac{1}{2} \nabla^\dalpha U D - \frac{1}{2} \nabla_\dbeta U M^{\dbeta \dalpha}
	- \frac{3i}{4} \nabla^\dalpha U A
\end{align}
These new derivatives are constructed so that when they act on a conformally primary
object, the result is conformally primary.

We are not the first to construct these objects. Kugo and Uehara,
in their treatment of conformal supergravity \cite{Kugo:1983mv}, constructed these operators
almost immediately out of the covariant derivatives, dubbing these the \textbf{u}-assisted
derivatives, where \textbf{u} denoted the compensator being used. Their motivation
seemed to be the desire for operators that would act on conformally primary superfields
to generate more conformally primary superfields. In that sense, these new
operators are special conformal \emph{invariant} rather than \emph{covariant}.

The purely undotted objects have a new algebra
\begin{align}
\{\CD_\beta, \CD_\alpha \} = \frac{1}{2} \left( \nabla^2 U + \nabla^\gamma U \nabla_\gamma U\right) M_{\beta \alpha}
	= \frac{1}{2} \frac{1}{X} \nabla^2 X M_{\beta \alpha}
	\equiv -4 \bar R M_{\beta \alpha}
\end{align}
Similarly,
\begin{align}
\{\CD^\dbeta, \CD^\dalpha \}
	= -4 R M^{\dbeta \dalpha}
\end{align}
where we have defined
\begin{align}
R \equiv -\frac{1}{8X} \bar \nabla^2 X, \;\;\;
\bar R \equiv -\frac{1}{8X} \nabla^2 X
\end{align}
From these definitions, $R$ possesses scaling and $U(1)_R$ weights $(\Delta, w) = (1,+2)$
and $\bar R$ the weights $(1,-2)$.
It is straightforward to show that in the limit where we gauge fix $X$ to unity,
these $R$'s become the $R$'s of Poincare supergravity. However, these versions are
more useful since they are also conformally invariant by nature of the
fact that the new covariant derivatives are themselves conformally invariant.
Furthermore, one may show that they are chiral with respect to the new derivatives:
\begin{align}
\CD^\dalpha R = 0, \;\;\; \CD_\alpha \bar R = 0.
\end{align}

It is straightforward to guess the form of the analogues of $G_c$ and $X_\alpha$.
Demanding that the definition of $G_c$ match when $X$ is fixed to unity (and also
be conformally invariant) gives
\begin{align}
G_{\alpha \dalpha} &= -\frac{1}{4} [\nabla_\alpha, \nabla_\dalpha] U + \frac{1}{4} \nabla_\alpha U \nabla_\dalpha U
	= \frac{1}{2} X^{1/2} [\nabla_\alpha, \nabla_\dalpha] X^{-1/2}
\end{align}
which is as we have defined it before.
Defining $X_\alpha$ as $\CD_\alpha R - \CD^\dalpha G_{\alpha \dalpha}$ leads to
\begin{align}
X_\alpha = \frac{3}{8} \bar\nabla^2 \nabla_\alpha U, \;\;\;
X^\dalpha = \frac{3}{8} \nabla^2 \nabla^\dalpha U
\end{align}
which is conformally invariant automatically.

We briefly pause to note the following features.
If $X = \Phi_0 \bPhi_0 e^{-K/3}$,
\[
X_\alpha = -\frac{1}{8} \bar\nabla^2 \nabla_\alpha K
	= -\frac{1}{8} (\BCD^2 - 8 R) \CD_\alpha K
\]
as in K\"ahler $U(1)$ supergravity.
Similarly, if $X=L$, then $R=0$ as in new minimal supergravity.

We next define the bosonic derivative $\CD_{\alpha \dalpha}$ by the anti-commutator
\begin{align}
\{\CD_\alpha, \CD^\dalpha \} \equiv -2i {\CD_\alpha}^\dalpha
	- \lambda G^{\beta \dalpha} M_{\beta \alpha}
	+ \lambda G_{\alpha \dbeta} M^{\dbeta \dalpha}
	+ 3i \lambda {G_\alpha}^\dalpha A
\end{align}
We have introduced into this definition a parameter $\lambda$ which
parametrizes how much of the various bosonic connections of $\CD_a$ is stored
in the additional ``curvatures'' on the right hand side.
$\lambda=1$ corresponds to the standard $U(1)$ supergravity of
Binetruy, Girardi, and Grimm \cite{bgg} and what is achieved by straightforwardly degauging
from conformal to Poincare supergravity \cite{Butter:2009cp}.
$\lambda=0$ corresponds to a redefinition
of that theory so that the $\alpha \dalpha$ curvatures are trivial. (This is
the choice made in \cite{superspace} and \cite{Buchbinder:1998qv}.) The
latter has the simplest-looking curvatures overall, but it introduces a
nonzero torsion $T_{cba}$ proportional to the dual of $G_a$, which leads to a bosonic Riemann
curvature tensor lacking the common symmetries and with an auxiliary superfield
hiding within the spin connection. For this reason $\lambda=0$ seems to be ill-suited for
component calculations; however, for the pure superfield manipulations we perform here,
it leads to a simpler algebra for the covariant derivatives. The two definitions
are completely equivalent, of course, and differ only in the definition of the bosonic
connections.

These definitions lead to
\begin{align}
{\CD_\alpha}^\dalpha \equiv& {\nabla_\alpha}^\dalpha
	- \frac{i}{2} \nabla_\alpha U \CD^\dalpha - \frac{i}{2} \nabla^\dalpha U \CD_\alpha
	- \frac{1}{2} {\nabla_\alpha}^\dalpha U D
	+ \left(+\frac{3}{8} [\nabla_\alpha,\nabla^\dalpha] U + \frac{3\lambda}{2} {G_\alpha}^\dalpha \right) A \eol
	& + \left(- \frac{i}{4} \nabla_\alpha \nabla_\dbeta U - \frac{i\lambda}{2} G_{\alpha \dbeta} \right)M^{\dbeta \dalpha}
	+ \left(- \frac{i}{4} \nabla^\dalpha \nabla^\beta U + \frac{i \lambda}{2} G^{\beta \dalpha} \right)M_{\beta \alpha}
\end{align}

The newly-defined curvatures are straightforward to work out.
For the bosonic-fermionic curvatures,
\begin{gather}
T_{\gamma (\beta \dbeta) \dalpha} = -2i \eps_{\gamma \beta} \eps_{\dbeta \dalpha} \bar R \\
T_{\gamma (\beta \dbeta) \alpha} =
	i \lambda G_{\gamma \dbeta} \eps_{\beta\alpha}
	-2i (1-\lambda)G_{\alpha \dbeta} \eps_{\gamma\beta} \\
F_{\beta (\alpha \dalpha)} = -\frac{3\lambda}{2} \CD_\beta G_{\alpha \dalpha} - \eps_{\beta \alpha} X_\dalpha \\
R_{\delta (\gamma \dgamma) \beta \alpha} =
	\sum_{\beta \alpha} \left[
		i \eps_{\delta \gamma} \CD_\beta G_{\alpha \dgamma}
		+ \frac{i\lambda}{2} \CD_\delta G_{\beta \dgamma} \eps_{\gamma \alpha}
		- i \eps_{\delta \beta} \eps_{\gamma \alpha} \CD_\dgamma \bar R
	\right] \\
R_{\delta (\gamma \dgamma) \dbeta \dalpha} =
	4i \eps_{\delta \gamma} W_{\dgamma \dbeta \dalpha}
	+ \sum_{\dbeta \dalpha} \eps_{\dgamma \dalpha} \left[
		\frac{i}{3} \eps_{\delta \gamma} \bar X_{\dbeta}
		+ \frac{i\lambda}{2} \CD_\delta G_{\gamma \dbeta}
	\right] 
\end{gather}
Note that these curvatures simplify a fair amount by choosing $\lambda=0$.

The bosonic torsions are
\begin{gather}
{T_{(\beta \dbeta)(\alpha \dalpha)}}^\gamma \CD_\gamma =
	-2\eps_{\dbeta \dalpha} W_{\beta \alpha \gamma} \CD^\gamma
	- \frac{1}{2} \eps_{\dbeta \dalpha} \CD_{\{\beta} R \;\CD_{\alpha \}}
	- \frac{1}{6} \eps_{\dbeta \dalpha} X_{\{\beta} \CD_{\alpha \}}
	- \frac{1}{2} \eps_{\beta \alpha} \CD_{\{\dbeta} G_{\dalpha\} \gamma} \CD^\gamma \\
T_{(\beta \dbeta)(\alpha \dalpha) \dgamma} \CD^\dgamma =
	-2 \eps_{\beta \alpha} W_{\dbeta \dalpha \dgamma} \CD^\dgamma
	+ \frac{1}{2} \eps_{\beta \alpha} \CD_{\{\dbeta} \bar R \;\CD_{\dalpha \}}
	+ \frac{1}{6} \eps_{\beta \alpha} \bar X_{\{\dbeta} \CD_{\dalpha \}}
	+ \frac{1}{2} \eps_{\dbeta \dalpha} \CD_{\{\beta} G_{\alpha\} \dgamma} \CD^\dgamma \\
{T_{(\beta \dbeta)(\alpha \dalpha)}}^c \CD_c =
	-2i (1-\lambda) G_{\beta \dalpha} \CD_{\alpha \dbeta}
	+2i (1-\lambda) G_{\alpha \dbeta} \CD_{\beta \dalpha}
\end{gather}
Note the last torsion vanishes for $\lambda=1$.

The part of the Riemann tensor acting on spinor indices is
\begin{align}
\frac{1}{2} R_{(\beta \dbeta)(\alpha \dalpha) \gamma \phi} M^{\phi\gamma } =&
	\eps_{\dbeta \dalpha} \sum_{\beta \alpha}\left(
	\frac{1}{2} \CD_{\beta} W_{\alpha \phi \gamma} M^{\gamma\phi}
	+ \frac{1}{12} \CD_\beta X^\gamma M_{\gamma \alpha}
	- \frac{1}{8} \BCD^2 R M_{\beta \alpha} + 2 R \bar R M_{\alpha \beta}
	\right) \eol
	& - \frac{1}{4} \eps_{\beta \alpha} \CD_{\{\dbeta} \CD_\gamma G_{\phi \dalpha\}} M^{\phi \gamma}
	- \frac{i\lambda}{2} \CD_{\beta \dbeta} {G^\phi}_{\dalpha} M_{\phi \alpha}
	+ \frac{i\lambda}{2} \CD_{\alpha \dalpha} {G^\phi}_{\dbeta} M_{\phi \beta} \eol
	&
	- \frac{\lambda^2}{2} G_{\beta \dalpha} {G^\phi}_{\dbeta} M_{\phi \alpha}
	+ \frac{\lambda^2}{2} G_{\alpha \dbeta} {G^\phi}_{\dalpha} M_{\phi \beta} \eol
	&+ \frac{1}{2} (\lambda^2-\lambda) \eps_{\dbeta \dalpha} G_{\phi \dphi} G^{\phi \dphi} M_{\beta \alpha}
\end{align}
The other half can be found by Hermitian conjugation.

The remaining $U(1)$ curvature is
\begin{align}
F_{(\beta \dbeta)(\alpha \dalpha)} = 
	-\frac{3\lambda}{2} \CD_{[(\beta \dbeta)} G_{(\alpha \dalpha)]}
	- \frac{i}{4} \eps_{\beta \alpha} \CD_{\{\dbeta} X_{\dalpha\}}
	- \frac{i}{4} \eps_{\dbeta \dalpha} \CD_{\{\beta} X_{\alpha\}}
\end{align}
Again note the simplifications which occur for the choice $\lambda=0$.

\subsubsection{Deformation}
The compensated derivatives (for $\lambda=0$) can be compactly written as
\begin{gather*}
\CD_\alpha \equiv \nabla_\alpha + \frac{1}{4} (\nabla^\beta U) \{S_\beta, Q_\alpha\}, \;\;\;
\CD^\dalpha \equiv \nabla^\dalpha + \frac{1}{4} (\nabla_\dbeta U) \{\bar S^\dbeta, \bar Q^\dalpha\} \\
\CD_{\alpha \dalpha} \equiv \frac{i}{2}\{\CD_\alpha, \CD_\dalpha\} 
\end{gather*}
provided we restrict them to only act on conformally primary objects. It
is in this form that it is easiest to demonstrate that if $\Psi$ is
primary, so is $\CD_\alpha \Psi$ where $\Psi$ possesses arbitrary weights
and Lorentz indices.

We have previously argued that to first order the spinor derivatives vary
(in Hermitian quantum gauge) as
$\delta \nabla_\alpha = [iV, \nabla_\alpha]$ and 
$\delta \nabla_\dalpha = [-iV, \nabla_\dalpha]$, where we had expanded
\[
V \equiv V^A \nabla_A + V^{\ul b} X_{\ul b}
\]
It follows then that the compensated spinor derivatives should vary as
\begin{align}
\delta \CD_\alpha &= [i V, \nabla_\alpha]
		+ \frac{1}{4} \left([iV, \nabla_\beta] U + \nabla_\beta \delta U\right) \{S_\beta, Q_\alpha\} \eol
	&= [i V, \CD_\alpha] + \frac{1}{4} \nabla^\beta (-i V U + \delta U) \{S_\beta, Q_\alpha\}
\end{align}
where we have substituted $\CD$ for $\nabla$ in the commutator. Note that
$(-i V U + \delta U)$ is conformally primary of dimension zero, and so we may
replace the $\nabla^\beta$ acting on it with $\CD^\beta$. Further simplifications
arise if we choose to expand $V$ in terms of the compensated derivative rather
than the covariant derivative:
\[
V = V^A \nabla_A + V^{\ul b} X_{\ul b} = V^{A'} \CD_A + V^{\ul b'} X_{\ul b}
\]
One may check that the $V'$'s are now conformally primary objects.
In particular, it is easy to show (by considering the variation of a chiral
superfield of vanishing weight for example) that
\begin{gather}
V'^a = V^a, \;\;\;
\tilde V'^\alpha = \tilde V^\alpha
\end{gather}
where $V'^\alpha \equiv -\frac{i}{8} \CD_\dphi V'^{\dphi \alpha} + \tilde V'^\alpha$.
Then provided we define a theory entirely in terms of $V^a$ and
$\tilde V^\alpha$, we can make use of these conformally invariant
derivatives when we calculate deformations of the quantum theory.

Henceforth we suppress the primes and trade the conformally covariant prepotentials
for the conformally invariant (or compensated) ones. One can show that
\begin{align}
V(D) &=  \frac{1}{2} \CD_b V^b + \frac{1}{2} \CD^\alpha V_\alpha
	+ \frac{1}{2} \CD_\dalpha V^\dalpha + \tilde V(D) \\
V(A) &= -\frac{1}{4} \Delta_b V^b + V^b G_b
	-\frac{3i}{4} \CD^\alpha V_\alpha + \frac{3i}{4} \CD_\dalpha V^\dalpha + \tilde V(A) \\
V(M)_{\beta \alpha} &=
	+ \frac{1}{2} \CD_{\{\beta} V_{\alpha\}}
	+ \frac{i}{8} \CD^\dphi \CD_{\{\beta} V_{\alpha \} \dphi}
	+ \frac{i}{2} V_{\{\alpha \dphi} {G_{\beta\}}}^\dphi
	+ \tilde V(M)_{\beta \alpha} \\
V(M)_{\dbeta \dalpha} &=
	+ \frac{1}{2} \CD_{\{\dbeta} V_{\dalpha\}}
	- \frac{i}{8} \CD^\phi \CD_{\{\dbeta} V_{\dalpha \} \phi}
	- \frac{i}{2} {V_{\{\dalpha}}^\phi G_{\dbeta\} \phi}
	+ \tilde V(M)_{\dbeta \dalpha}
\end{align}
Note the forms are quite similar to what we had in \eqref{eq_V's}, except for
the appearance of the new superfield $G_b$. We have also introduced the
conformally invariant operator $\Delta_{\alpha \dalpha} = -\frac{1}{2} [\CD_\alpha, \BCD_\dalpha]$.

Since $U$ obeys $\CD_A U = 0$, it follows that
\begin{align}
\delta \CD_\alpha
	= [i V, \CD_\alpha] + \frac{1}{4} \nabla^\beta (-2i V(D) + \delta U) \{S_\beta, Q_\alpha\}
\end{align}
from which we may derive the variations of each of the spinor connections.
We find
\begin{align}
H_{\alpha \beta} &= i \CD_\alpha V_\beta - i V^c  T_{\alpha c \beta}
	-i V(M)_{\alpha \beta} + \frac{i}{2} V(D) \eps_{\alpha \beta} + V(A) \eps_{\alpha \beta} \eol
H_{\alpha \dbeta} &= i \CD_\alpha V_\dbeta - i V^c T_{\alpha c \dbeta} \eol
H_{\alpha (\beta \dbeta)} &= i \CD_\alpha V_{\beta \dbeta} + 4 V_\dbeta \eps_{\alpha \beta} \eol
\Lambda_\alpha &= \frac{1}{2} \CD_\alpha (\delta U) \eol
\omega_\alpha &= i \CD_\alpha V(A) - i V^b F_{\alpha b} - \frac{3}{2} \CD_\alpha V(D) - \frac{3i}{4} \CD_\alpha \delta U \eol
\Omega_\alpha(M) &= i \CD_\alpha V(M) + 4i \bar R V^\beta M_{\beta \alpha}
	- i V^b R_{\alpha b}(M) - i \CD^\beta V(D) M_{\beta \alpha} + \frac{1}{2} \CD^\beta \delta U M_{\beta \alpha}
\end{align}
and for their conjugates
\begin{align}
H_{\dalpha \beta} &= -i \CD_\dalpha V_\beta + i V^c T_{\dalpha c \beta} \eol
H_{\dalpha \dbeta} &= -i \CD_\dalpha V_\dbeta + i V^c  T_{\dalpha c \dbeta}
	+i V(M)_{\dalpha \dbeta} + \frac{i}{2} V(D) \eps_{\dalpha \dbeta} - V(A) \eps_{\dalpha \dbeta} \eol
H_{\dalpha (\beta \dbeta)} &= -i \CD_\dalpha V_{\beta \dbeta} + 4 V_\beta \eps_{\dalpha \dbeta} \eol
\Lambda_\dalpha &= \frac{1}{2} \CD_\dalpha (\delta U) \eol
\omega_\dalpha &= -i \CD_\dalpha V(A) + i V^b F_{\dalpha b}
	- \frac{3}{2} \CD_\dalpha V(D) + \frac{3i}{4} \CD_\dalpha \delta U \eol
\Omega^\dalpha(M) &= -i \CD^\dalpha V(M) - 4i R V_\dbeta M^{\dbeta \dalpha}
	+ i V_b R^{\dalpha b}(M) + i \CD_\dbeta V(D) M^{\dbeta \dalpha} + \frac{1}{2} \CD_\dbeta \delta U M^{\dbeta \dalpha}
\end{align}

The variation of the bosonic derivatives is straightforward to work out from
the above results. Using these, one may for example work out the variations of
the superfields $G_{\alpha \dalpha}$ and $R$ in the language of these
compensated derivatives. For $R$, it is actually easier to work in the original
theory at first. Recall the chiral variation of an arbitrary superfield $\Psi$ can
be defined by
\begin{align}
\delta_c \Psi = \delta \Psi + i V \Psi
\end{align}
which generalizes the case where $\Psi$ is itself chiral.
Then the chiral variation of $R$ is
\begin{align}
\delta_c R &= -\frac{1}{8X} \bar\nabla^2 \left(X \delta_c U\right) + \frac{1}{8X} \delta_c U \bar \nabla^2 X
	= -\frac{1}{8} \BCD^2 \delta_c U
\end{align}
Similarly, the chiral variation of $X_\alpha$ is
\begin{align}
\delta_c X_\alpha &= \frac{3}{8} \bar\nabla^2 \left(\nabla_\alpha \delta U + 2i V \nabla_\alpha U - i \nabla_\alpha (V U)\right) \eol
	&= \frac{3}{8} (\BCD^2 - 8 R) \left(\CD_\alpha \delta U + 2i Z_\alpha
	- 2i \CD_\alpha V(D)\right)
\end{align}
where
\begin{align}
Z_\alpha \equiv&\, V\nabla_\alpha U \eol
	=& - \frac{1}{2} (\CD^2 - 12 \bar R) V_\alpha
	+ \frac{1}{2} \CD^\dbeta \CD_\alpha V_\dbeta \eol
	& - \frac{1}{6} \CD_\alpha \CD_{\beta \dbeta} V^{\dbeta \beta}
	+ \frac{i}{3} \CD_\alpha (G_{\beta \dbeta} V^{\dbeta \beta})
	+ \frac{i}{24} (\CD^2 - 12 \bar R) \CD^\dbeta V_{\dbeta \alpha} \eol
	& + \CD^\dbeta (\bar R V_{\alpha \dbeta})
	+ \frac{i}{3} X^\dbeta V_{\alpha \dbeta}
\end{align}

Calculating $\delta G_{\alpha \dalpha}$ is a bit more difficult
since its definition in terms of $X$ necessarily involves both dotted and undotted
spinor derivatives in a symmetric fashion.
The most straightforward way to proceed seems to be to work out
its variation by calculating the variation of the torsion component
$\delta T_{\gamma b \alpha}$. This gives the following rather complicated
expression:
\begin{align*}
\delta G_{\alpha \dalpha} =& -\frac{1}{4} [\CD_\alpha, \CD_\dalpha] \tilde \delta U - {H_{\alpha \dalpha}}^b G_b
	- i V^\beta \CD_\beta G_{\alpha \dalpha} -i V^\dbeta \CD_\dbeta G_{\alpha \dalpha} \\
	& - \frac{1}{2} \Delta_{\alpha \dalpha} \Delta_b V^b
	-\frac{1}{2} \CD_{\alpha \dalpha} \CD_b V^b - \frac{1}{32} (\CD (\BCD^2 - 8R) \CD + \hc) V_{\alpha \dalpha} \\
	& + \frac{1}{2} (\CD^\gamma {V_{\dalpha}}^\beta) W_{\gamma \beta \alpha}
	+\frac{1}{2} (\CD^\dgamma {V_{\alpha}}^\dbeta) W_{\dgamma \dbeta \dalpha}
	-\frac{1}{2} \Delta_{\alpha \dalpha} (V^b G_b) \\
	& + \frac{1}{8} \CD^\beta V_{\beta \dalpha} \CD_\alpha R
	+ \frac{1}{8} \CD_\alpha V_{\beta \dalpha} \CD^\beta R
	+ \frac{1}{6} \CD_{\{\beta} V_{\alpha\} \dalpha} X^\beta \\
	& - \frac{1}{8} \CD^\dbeta V_{\dbeta \alpha} \CD_\dalpha R
	- \frac{1}{8} \CD_\dalpha V_{\dbeta \alpha} \CD^\dbeta R
	- \frac{1}{6} \CD_{\{\dbeta} V_{\dalpha\} \alpha} X^\dbeta \\
	& - \bar R R V_{\alpha \dalpha} - \frac{1}{4} V_{\alpha \dalpha} \CD^\beta X_\beta \\
	& -\frac{1}{2} (\Delta_{\alpha \dalpha} V^b)  G_b
	-\frac{1}{2} \Delta_b (V^b G_{\alpha \dalpha}) +\frac{1}{2} (\Delta_b V^b) G_{\alpha \dalpha} \\
	& + \frac{i}{4} V_{\beta \dalpha} \CD^{\dbeta \beta} G_{\alpha \dbeta}
	- \frac{i}{4} V_{\alpha \dbeta} \CD^{\dbeta \beta} G_{\beta \dalpha}
\end{align*}
where we have defined
\begin{align}\label{eq_deftdU}
\tilde \delta U \equiv \delta U + i \CD^\beta V_\beta - i \CD_\dbeta V^\dbeta + \Delta_b V^b.
\end{align}
For the linear compensator model, $\tilde \delta U = L^{-1} \mathcal L$,
but for the generic chiral model
\begin{align}
\tilde \delta U = -\frac{1}{3} \left(
	\K_i \eta^i + \K_\bj \bar\eta^\bj
	- 2 \Delta_b V^b - 4 V^b G_b + V^b \K_b
	\right)
\end{align}

The expression for $\delta G_{\alpha \dalpha}$ involves a combination of the supergravity potentials
that has been succinctly combined into $H_a{}^b$, which is the deformation of the bosonic vierbein.
It can be calculated from
\[
\delta \CD_a = -H_a{}^B \CD_B - H_a{}^{\ul b} X_{\ul b},
\]
the left hand side of which can itself be calculated easily from $\delta\CD_\alpha$
and $\delta \CD_\dalpha$. The reason for collecting these terms in this
way is that we will eventually find they cancel out.

Rearranging a number of terms leads to
\begin{align}\label{eq_dG0}
\delta G_{\alpha \dalpha} =& \frac{1}{2} \Delta_{\alpha \dalpha} \tilde \delta U - {H_{\alpha \dalpha}}^b G_b
	- i V^\beta \CD_\beta G_{\alpha \dalpha} -i V^\dbeta \CD_\dbeta G_{\alpha \dalpha} \eol
	& - \frac{1}{2} \Delta_{\alpha \dalpha} \Delta_b V^b
	- \frac{1}{2} \CD_{\alpha \dalpha} \CD_b V^b
	- \frac{1}{32} \{\CD^2,\BCD^2\} V_{\alpha \dalpha}
	+ \frac{1}{2} \Box V_{\alpha \dalpha} \eol
	& + \frac{1}{2} (R \CD^2 + \bar R \BCD^2) V_{\alpha \dalpha}
	+ (\CD^\gamma {V_{\dalpha}}^\beta) W_{\gamma \beta \alpha}
	+ (\CD^\dgamma {V_{\alpha}}^\dbeta) \bar W_{\dgamma \dbeta \dalpha} \eol
	& - G^b \Delta_b V_{\alpha \dalpha}
	- (\Delta_{\alpha \dalpha} V^b) G_b
	- \Delta_b (V^b G_{\alpha \dalpha})
	+ (\Delta_b V^b) G_{\alpha \dalpha}
	+ \frac{1}{2} V^b \Delta_{[b} G_{(\alpha \dalpha)]} \eol
	& + \frac{1}{2} \CD^\beta V_{\beta \dalpha} (\CD_\alpha R - \frac{1}{3} X_\alpha)
	+ \frac{1}{12} \CD^\beta V_{\alpha \dalpha} X_\beta \eol
	& - \frac{1}{2} \CD^\dbeta V_{\dbeta \alpha} (\CD_\dalpha \bar R - \frac{1}{3} X_\dalpha)
	+ \frac{1}{12} \CD_\dbeta V_{\alpha \dalpha} X^\dbeta \eol
	& - \bar R R V_{\alpha \dalpha} - \frac{1}{8} V_{\alpha \dalpha} (\CD^\beta X_\beta + \hc)
	+ \frac{i}{4} V_{\beta \dalpha} \CD^{\dbeta \beta} G_{\alpha \dbeta} 
	- \frac{i}{4} V_{\alpha \dbeta} \CD^{\dbeta \beta} G_{\beta \dalpha}
\end{align}

\subsection{Proceeding to second order}
We would like to proceed to second order so that we can perform one-loop
calculations. The immediate difficulty we face is that we solved our constraints
only to first order. For example, $\tilde V(A)$
might also involve some second order object of the form $V^a \mathcal O_a{}^b V_b$
where $\mathcal O_a{}^b$ is some conformally invariant operator. Then in analyzing the
variations of the $\mathcal W$'s, we should have worked to second order
in $V^a$ to find out if any such object exists.

There are two approaches one could take at this point. One would be to return
to the original analysis and redo it to second order and determine what
modifications are necessary. The second approach is to use our ability to
take first order variations and to vary to first order the first order action that
we already have -- thereby bootstrapping to second order. This is possible since
our first order solution was not dependent on any specific origin point on
the constraint surface of conformal supergravity; it merely required that we
remain \emph{somewhere} on that surface.

This latter approach is the one we will take. The main difficulty 
is figuring out how to vary the quantum superfields
$V^a$ and $\Sigma^r$. On the one hand, varying these
only shifts the action by a term proportional to the equations of motion,
so it's not an immediate issue if we choose to work on shell. On the other,
if there is some sort of natural variation of these objects, then we
can possibly simplify the second-order action without the need to apply
the equations of motion.

We begin by considering a primary chiral superfield of vanishing weights.
In this way its variation can be defined solely in terms of $V^a$ and $\tilde V^\alpha$.
Then varying $\Phi$ in the most natural way amounts to
\begin{align}
\Phi' = e^{-i V} (\Phi + \eta)
	= \Phi - i V \Phi + \eta - \frac{1}{2} V^2 \Phi - i V \eta + \mathcal O(V^3)
\end{align}
where we have stopped the expansion at second order.
Demanding that the second order terms agree with the first order
variation of the first order terms gives
\begin{gather}
\delta \eta = -i V \eta, \;\;\;
\delta (V\Phi) = -i V^2 \Phi
\end{gather}
(In the calculation one must include an additional factor of 2 since
the second variation is generated from half of the first order variation squared.)
The first is a perfectly sensible definition (it amounts to $\delta_c \eta=0$)
and the second implies for the variations of $V^a$ and $\Sigma^r$
\begin{align}
\delta V_{\alpha \dalpha} &= -8 V_\alpha V_\dalpha + i V^\beta \CD_\beta V_{\alpha \dalpha}
	- i V_\dbeta \BCD^\dbeta V_{\alpha \dalpha}
	+ V^b H_{b (\alpha \dalpha)} \eol
\delta \Sigma &= i V^\alpha \CD_\alpha \Sigma - i V_\dalpha \CD^\dalpha \Sigma
	+ 2i V^\alpha V^b F_{b \alpha}
	+ 2i V^\dalpha V^b F_{b \dalpha}
	- V^a \Delta_a \Sigma \eol
	&\;\;\;
	+ V^{\dalpha \alpha} \left(
		- \frac{1}{2} \CD_\alpha V^b F_{b \dalpha}
		+ \frac{1}{2} \CD_\dalpha V^b F_{b \alpha}
		- \frac{1}{4} V^b \CD_\alpha F_{b \dalpha}
		+ \frac{1}{4} V^b \CD_\dalpha F_{b \alpha}
	\right)
\end{align}
In the last equation we have suppressed the $r$ index to simplify notation.

Note that $\delta V_a \ni V^b H_b{}^a$ and $\delta G_a = -H_a{}^b G_b$
and so there will be no $H_b{}^a$ in terms like $\delta (V^a G_a)$.
We will similarly identify the combination $H_a{}^b$ in the
variation of $\K_a$ and $\mathcal Y_a$ so that this cancellation occurs
for these terms as well.

\subsubsection{Variation of the $\eta$ term}
Beginning with
\begin{align}
\delta_\eta S = \int \chE\, \eta^i \left( X \CP \K_i + P_i + \mathcal Y_i \right) + \hc
\end{align}
we consider the effect of a second variation. Given the presence of $\eta^i$,
it is most sensible to work in quantum chiral gauge where $\eta^i$ has no further
variation.

Taking the superpotential term, one finds simply
\begin{align}
\delta \delta_\eta S \ni
	\int \chE\, \eta^i \eta^j P_{ij}
\end{align}
The gauge field term is a bit more complicated:
\begin{align}
\delta \delta_\eta S \ni
	\int \chE\, \eta^i \left(\frac{1}{4} \eta^j f_{rs,ij} W^{\alpha r} W_{\alpha}^s +
		\frac{1}{2} f_{rs,i} W^{\alpha r} \delta_c W_{\alpha}^s \right)
\end{align}
Plugging in $\delta_c W_{\alpha}^s$ gives
\begin{align}
\delta \delta_\eta S \ni
	\frac{1}{4} \int \chE\, \eta^i \eta^j f_{rs,ij} W^{\alpha r} W_{\alpha}^s
	+ \frac{1}{2} \int E\, \eta^i f_{rs,i} W^{\alpha r} \left(i \nabla_\alpha \Sigma^s
	+ V_{\alpha \dalpha} \bar W^{\dalpha s} \right)
\end{align}

The term involving $X$ and $\K_i$ is the most difficult to deal with.
We rewrite it as a full superspace integral and then take the chiral quantum variation\footnotemark
\begin{align}
\delta \delta_\eta S \ni \delta_c X \eta^i \K_i
	+ X \eta^i \K_{i \bj} \left(\bar\eta^\bj + 2i V^B X_B \bPhi^\bj \right)
	+ X \eta^i \K_{i j} \eta^j
\end{align}
The last term we will consider in tandem with $P_{ij}$. The second term can be simplified
by noting that when $X_B = D$ or $A$, the result simplifies. First note
\begin{align}
D \K_i = - \Delta_i \K_i &= +\K_{ij} \Delta_j \Phi^j + \K_{i \bj} \Delta_\bj \bPhi^\bj \\
\frac{3i}{2} A \K_i = \Delta_i \K_i &= -\K_{ij} \Delta_j \Phi^j + \K_{i \bj} \Delta_\bj \bPhi^\bj
\end{align}
which together imply
\begin{align}
0 = \K_{i\bj} \Delta_{\bj} \bPhi^\bj.
\end{align}
This gives
\begin{align}
\K_{i \bj} \eta^i \bar\eta^\bj
+ 2i \eta^i \K_{i \bj} \left(V^b \CD_b + \Sigma^r X_r\right) \bPhi^\bj
 + 2i V_\dalpha \CD^\dalpha (\eta^i \K_i)
\end{align}

\footnotetext{We have written this and many subsequent D-terms without an
overall $\int E$ or with the brackets $[ \,\,\,]_D$ to keep the
formulae from growing cluttered.}

Next we observe that $\delta_c X$ is equivalent to
\begin{align}
\delta_c X = X \delta U + i V X 
	= X \tilde \delta U - i X \CD^\beta V_\beta + i X \CD_\dbeta V^\dbeta - X \Delta_b V^b + 2 i V(D) X
\end{align}
where we have used \eqref{eq_deftdU} again. Plugging this in and using
several integrations by parts, we can show that the total variation of
this term is
\begin{align}
\delta \delta_\eta S \ni
	X \left(
	i \CD_b V^b  K_i \eta^i - \Delta_b V^b K_i \eta^i
	+ 2i \eta^i K_{i \bj} (V+\Sigma) \bPhi^\bj
	+ \tilde\delta U K_i \eta^i
	+ K_{i j} \eta^i \eta^j
	+ K_{i \bj} \eta^i \bar\eta^\bj
	\right)
\end{align}
The combination $(V+\Sigma)$ is shorthand for $(V^b \CD_b + \Sigma^r X_r)$.
Note that the terms involving $V^\alpha$ and $V_\dalpha$ have dropped out.
We can simplify this expression by combining the first two terms and then
integrating by parts. The result is
\begin{align}
\delta \delta_\eta S \ni
	X \left(
	- \frac{1}{2} V^{\dalpha \alpha} \CD_\alpha (K_{i \bj} \eta^i) \CD_\dalpha \bPhi^\bj
	+ 2i \eta^i K_{i \bj} \Sigma \bPhi^\bj
	+ \tilde\delta U K_i \eta^i
	+ K_{i j} \eta^i \eta^j
	+ K_{i \bj} \eta^i \bar\eta^\bj
	\right)
\end{align}
Combining this with everything else yields
\begin{align}
\delta \delta_\eta S
	&= \left[\eta^i (\CP X K_{i j} + P_{ij} + \mathcal Y_{ij})\eta^j\right]_F + \hc \eol 
	& \,\,\, + \left[
	X \eta^i K_{i \bj} \bar \eta^\bj
	+ X \tilde\delta U \eta^i K_i
	+ \eta^i \left(X K_{i,r} + \mathcal Y_{i,r} \right) \Sigma^r 
	+ X V^a K_{a, i} \eta^i
	+ V^a \mathcal Y_{a, i} \eta^i \right]_D + \hc \eol
\end{align}
where we have defined
\begin{align}
\K_{i, r} &\equiv 2i \K_{i \bj} X_r \bPhi^\bj \\
\mathcal Y_{i, r} & \equiv \frac{i}{2} f_{rs,i} W^{\alpha s} \CD_\alpha
	= \frac{i}{2} G_{rs,i} W^{\alpha s} \CD_\alpha \\
\K_{\alpha \dalpha, i} \eta^i &\equiv -\CD_\dalpha \bPhi^\bj \CD_\alpha \left(K_{i \bj} \eta^i\right) \\
\mathcal Y_{\alpha \dalpha, i} &\equiv -f_{rs,i} W_\alpha^r \bar W_\dalpha^s
	= -G_{rs,i} W_\alpha^r \bar W_\dalpha^s
\end{align}

\subsubsection{Variation of the $\Sigma$ term}
The $\Sigma$ term is
\[\int E \, \Sigma^r (\mathcal Y_r + X K_r ) \]
where we recall
\begin{gather*}
\mathcal Y_r \equiv -\frac{i}{2} \nabla^\alpha \left(f_{rs} W_\alpha{}^s \right) + \hc \\
K_r \equiv -i K_i X_r \Phi^i + i K_{\bj} X_r \bPhi^\bj
\end{gather*}

The variation of the first term is given by using the formula
\begin{align}
\delta (\nabla^\alpha \Phi_\alpha) =
	& -i \nabla^\beta (V_\beta \nabla^\alpha \Phi_\alpha)
	+i \nabla_\dbeta (V^\dbeta \nabla^\alpha \Phi_\alpha)
	- \Delta_b (V^b \nabla^\alpha \Phi_\alpha)
	+ 2 V^{\dalpha \alpha} \bar W_\dalpha \Phi_\alpha \eol
	& - i \Sigma \nabla^\alpha \Phi_\alpha - 2i (\nabla^\alpha \Sigma^r) X_r \Phi_\alpha 
	+ \frac{1}{4} \nabla_\dalpha \nabla^2 (V^{\dalpha \alpha} \Phi_\alpha)
	+ \nabla^\alpha (\delta_c \Phi_\alpha)
\end{align}
where $\Phi_\alpha$ is an arbitrary chiral spinor superfield. This is written
in terms of the old $V_\beta$ and $V^\dbeta$. Exchanging for the new conformally
invariant ones gives
\begin{align}
\delta (\nabla^\alpha \Phi_\alpha) =
	& -i \CD^\beta (V_\beta \nabla^\alpha \Phi_\alpha)
	+i \CD_\dbeta (V^\dbeta \nabla^\alpha \Phi_\alpha)
	- \Delta_b (V^b \nabla^\alpha \Phi_\alpha)
	+ 2 V^{\dalpha \alpha} \bar W_\dalpha \Phi_\alpha \eol
	& - i \Sigma \nabla^\alpha \Phi_\alpha - 2i (\CD^\alpha \Sigma^r) X_r \Phi_\alpha
	+ \frac{1}{4} \nabla_\dalpha \nabla^2 (V^{\dalpha \alpha} \Phi_\alpha)
	+ \CD^\alpha (\delta_c \Phi_\alpha)
\end{align}
In this formula, we have mixed conventions with $\nabla$'s and $\CD$'s appearing
in the same expression. Every isolated $\nabla_\alpha$ (or $\nabla_\dalpha$) here
is equivalent to $\CD_\alpha$ (or $\CD_\dalpha$), while $\nabla^2$ is equivalent to
$\CD^2 - 8 \bar R$. $\Delta_b$ is in terms of $\CD$
and this will remain the case for the rest of this work.

Applying this formula to $\mathcal Y_r$ gives
\begin{align}
\delta \mathcal Y_r =&
	- i \CD^\beta (V_\beta \mathcal Y_r)
	+ i \CD_\dbeta (V^\dbeta \mathcal Y_r)
	- \Delta_b (V^b \mathcal Y_r) \eol
	& + i V^{\dalpha \alpha} W_{\dalpha}{}^s W_{\alpha}{}^u f_{sr}{}^t f_{tu}
	+ i V_{\alpha \dalpha} W^{\alpha}{}^s \bar W^{\dalpha}{}^u f_{sr}{}^t \bar f_{tu} \eol
	& + \frac{i}{8} \nabla^\alpha \bar\nabla^2 (V_{\alpha \dalpha} G_{rs} \bar W^{\dalpha r})
	- \frac{i}{8} \nabla_\dalpha \nabla^2 (V^{\dalpha \alpha} G_{rs} W_{\alpha}{}^s) \eol
	& + i \Sigma^s f_{sr}{}^t \mathcal Y_t
	+ \CD^\alpha \Sigma^s f_{sr}{}^t W_{\alpha}{}^u f_{tu}
	- \BCD_\dalpha \Sigma^s f_{sr}{}^t \bar W^\dalpha{}^{u} \bar f_{tu} \eol
	& - \frac{1}{8} \nabla^\alpha \left(f_{rs} \bar\nabla^2 \nabla_\alpha \Sigma^s \right)
	- \frac{1}{8} \nabla_\dalpha \left(\bar f_{rs} \nabla^2 \nabla^\dalpha \Sigma^s \right) \eol
	& - \frac{i}{2} \CD^\alpha (\eta^i f_{rs,i} W_\alpha{}^s)
	- \frac{i}{2} \CD_\dalpha (\bar\eta^\bj f_{rs,\bj} \bar W^\dalpha{}^s)
\end{align}
Including the variation of $\Sigma$ and integrating by parts gives
\begin{align}
\delta(\Sigma^r \mathcal Y_r) =&
	\left(2i V^\alpha \CD_\alpha \Sigma^r - 2i V_\dalpha \CD^\dalpha \Sigma^r
	+ 2i V^\alpha V^b F_{b \alpha}
	+ 2i V^\dalpha V^b F_{b \dalpha} \right) \mathcal Y_r \eol
	& + V^{\dalpha \alpha} \left(
		- \frac{1}{2} \CD_\alpha V^b F_{b \dalpha}
		+ \frac{1}{2} \CD_\dalpha V^b F_{b \alpha}
		- \frac{1}{4} V^b \CD_\alpha F_{b \dalpha}
		+ \frac{1}{4} V^b \CD_\dalpha F_{b \alpha}
	\right) \mathcal Y_r \eol
	& - 2 V^a (\Delta_a \Sigma^r) \mathcal Y_r \eol
	& + \Sigma^r V^a \mathcal Y_{a, r}
	+ \frac{i}{8} \Sigma^r \nabla^\alpha \bar\nabla^2 (V_{\alpha \dalpha} G_{rs} \bar W^{\dalpha r})
	- \frac{i}{8} \Sigma^r \nabla_\dalpha \nabla^2 (V^{\dalpha \alpha} G_{rs} W_{\alpha}{}^s) \eol
	& + \Sigma^r \CD^\alpha \Sigma^s f_{sr}{}^t W_{\alpha}{}^u f_{tu}
	- \Sigma^r \BCD_\dalpha \Sigma^s f_{sr}{}^t \bar W^\dalpha{}^{u} \bar f_{tu} \eol
	& - \frac{1}{8} \Sigma^r \nabla^\alpha \left(f_{rs} \bar\nabla^2 \nabla_\alpha \Sigma^s \right)
	- \frac{1}{8} \Sigma^r \nabla_\dalpha \left(\bar f_{rs} \nabla^2 \nabla^\dalpha \Sigma^s \right) \eol
	& + \eta^i \mathcal Y_{i r} \Sigma^r
	+ \bar\eta^\bj \mathcal Y_{\bj r} \Sigma^r
\end{align}
where we have defined
\begin{align}
\mathcal Y_{\alpha \dalpha, r} &\equiv -2i \left(
	\bar W_\dalpha^s W_\alpha^u f_{sr}{}^t f_{tu}
	+ W_\alpha^s \bar W_\dalpha^u f_{sr}{}^t \bar f_{tu}
	\right)
\end{align}

Varying $\K_r$ gives
\begin{align}
\delta \K_r
	&= -i V^\beta \CD_\beta \K_r + i V_\dbeta \CD^\dbeta \K_r
	+  \eta^i \K_{i r} + \bar\eta^\bj \K_{\bj r} \eol
	& \;\;\;\;\;
	+ 2 \K_{i \bj} (V+\Sigma) \Phi^i X_r \bPhi^\bj
	+ 2 \K_{i \bj} X_r \Phi^i (V + \Sigma) \bPhi^\bj
\end{align}
where again
\[
V + \Sigma \equiv V^b \CD_b + \Sigma^r X_r
\]

Including the variation of $X$ and $\Sigma^r$ gives
\begin{align}
X^{-1} \delta (\Sigma^r X \K_r)
	=& \left(2i V^\alpha \CD_\alpha \Sigma^r - 2i V_\dalpha \CD^\dalpha \Sigma^r
	+ 2i V^\alpha V^b F_{b \alpha}
	+ 2i V^\dalpha V^b F_{b \dalpha} \right) \K_r \eol
	& + V^{\dalpha \alpha} \left(
		- \frac{1}{2} \CD_\alpha V^b F_{b \dalpha}
		+ \frac{1}{2} \CD_\dalpha V^b F_{b \alpha}
		- \frac{1}{4} V^b \CD_\alpha F_{b \dalpha}
		+ \frac{1}{4} V^b \CD_\dalpha F_{b \alpha}
	\right) \K_r \eol
	& - V^a \Delta_a \Sigma^r \K_r
	- (\Delta_b V^b) \Sigma^r \K_r
	+ \tilde \delta U \Sigma^r \K_r \eol
	& 
	+ 4 \K_{i \bj} \Sigma \Phi^i \Sigma \bPhi^\bj
	+ 2 \K_{i \bj} V \Phi^i \Sigma \bPhi^\bj
	+ 2 \K_{i \bj} \Sigma \Phi^i V \bPhi^\bj  \eol
	&
	+ \Sigma^r \K_{i, r} \eta^i + \Sigma^r \K_{\bj, r} \bar\eta^\bj
\end{align}


\subsubsection{Variation of the $V^a$ term}
The $V^a$ term is
\begin{align}
\left[V^b (-4 X G_b + X K_b + \mathcal Y_b) \right]_D
\end{align}
We require the variations of $G_{\alpha \dalpha}$, 
$K_{\alpha \dalpha}$, and $\mathcal Y_{\alpha \dalpha}$ in order
to continue.

The variation of $G_{\alpha \dalpha}$ contains the graviton kinetic term.
We have already worked this out in \eqref{eq_dG0}, but we rewrite it here
in the compact and useful form
\begin{align}\label{eq_dG}
X^{-1} \delta (X G_{\alpha \dalpha})
	=&
	\,\tilde \delta U G_{\alpha \dalpha}
	+ \frac{1}{2} \Delta_{\alpha \dalpha} \tilde \delta U - {H_{\alpha \dalpha}}^b G_b
	- i \CD^\beta (V_\beta G_{\alpha \dalpha}) + i \CD_\dbeta (V^\dbeta G_{\alpha \dalpha}) \eol
	& - \frac{1}{2} \Delta_{\alpha \dalpha} \Delta_b V^b
	- \frac{1}{2} \CD_{\alpha \dalpha} \CD_b V^b
	- \frac{1}{32} \{\CD^2,\BCD^2\} V_{\alpha \dalpha}
	+ \frac{1}{2} \Box_V V_{\alpha \dalpha} \eol
	& - G_{\alpha \dalpha} \Delta_b V^b - \Delta_{\alpha \dalpha} (V^b G_b)
	+ \frac{1}{2} \CD^\beta (R \CD_\beta V_{\alpha \dalpha})
	+ \frac{1}{2} \CD_\dbeta (\bar R \CD^\dbeta V_{\alpha \dalpha})  \eol
	& - \frac{1}{2} \CD_\alpha V_\dalpha{}^\beta X_\beta
	+ \frac{1}{2} \CD_\dalpha V_\alpha{}^\dbeta X_\dbeta
	- \frac{1}{2} V_b \CD_c G_d \eps_{dcba} \sigma^a_{\alpha \dalpha} \eol
	& - \frac{1}{8} V^{\dalpha \alpha} (\CD^2 R + \BCD^2 \bar R)
	- R \bar R V_{\alpha \dalpha}
	+ \frac{1}{2} V^{\dalpha \alpha} \Delta_b G^b
	+ \frac{1}{2} V^b \Delta_{[(\alpha \dalpha)} G_{b]}
\end{align}
where we have defined
\begin{align}
\Box_V V_{\alpha \dalpha} \equiv
	\Box V_{\alpha \dalpha}
	- \frac{1}{2} \CD^{[\beta} (G_{\beta \dbeta} \CD^{\dbeta]} V_{\alpha \dalpha})
	+ \frac{1}{2} \CD^\gamma V^{\dbeta \beta} W_{\gamma (\beta \dbeta) (\alpha \dalpha)}
	+ \frac{1}{2} \CD^\dgamma V^{\dbeta \beta} W_{\dgamma (\beta \dbeta) (\alpha \dalpha)}
\end{align}
$W_{\gamma b a}$ and its conjugate are defined by
\begin{align}
R_{\dot\delta (\gamma \dgamma) b a} &= 2i \eps_{\dot \delta \dgamma} W_{\gamma b a} \eol
R_{\delta (\gamma \dgamma) b a} &= 2i \eps_{\delta \gamma} W_{\dgamma b a}
\end{align}

The variation we need is
\begin{align}
X^{-1} \delta (-4 X V^a G_a)
	=
	& - 8i V^\beta \CD_\beta V^a G_a
	+ 8i V_\dbeta \CD^\dbeta V^a G_a
	- 16 V^\alpha V^\dalpha G_{\alpha \dalpha} \eol
	& -4 \tilde \delta U (V^a G_a)
	-2 \Delta_a V^a \tilde \delta U \eol
	& + 2 (\Delta_b V^b)^2
	- 2 (\CD_b V^b)^2
	- \frac{1}{8} \CD^2 V^{\dalpha \alpha} \BCD^2 V_{\alpha \dalpha}
	+ V^{\dalpha \alpha} \Box_V V_{\alpha \dalpha} \eol
	& + 8 V^a G_a \Delta_b V^b \eol
	& + V^{\dalpha \alpha} \CD^\beta (R \CD_\beta V_{\alpha \dalpha})
	+ V^{\dalpha \alpha} \CD_\dbeta (\bar R \CD^\dbeta V_{\alpha \dalpha})  \eol
	& - V^{\dalpha \alpha} \CD_\alpha V_\dalpha{}^\beta X_\beta
	+ V^{\dalpha \alpha} \CD_\dalpha V_\alpha{}^\dbeta X_\dbeta \eol
	& - \frac{1}{4} V^{\dalpha \alpha} V_{\alpha \dalpha} (\CD^2 R + \BCD^2 \bar R)
	- 2 R \bar R V^{\dalpha \alpha} V_{\alpha \dalpha}
	+ V^{\dalpha \alpha} V_{\alpha \dalpha} \Delta_b G^b
\end{align}
Note that the combination $H_a{}^b$ cancels out of the expression.

Turning to the variation of the matter term, we begin by noting that
$\K_{\alpha \dalpha}$ may be written a number of equivalent ways
\begin{align}
\K_{\alpha \dalpha} &= K_{i \bj} \nabla_\alpha \Phi^i \nabla_\dalpha \bPhi^\bj
	= \nabla_\alpha \Phi^i \nabla_\dalpha K_i
	= \nabla_\alpha K_\bj \nabla_\dalpha \bPhi^\bj \eol
	&= K_{i \bj} \CD_\alpha \Phi^i \CD_\dalpha \bPhi^\bj
	= \CD_\alpha \Phi^i \CD_\dalpha K_i
	= \CD_\alpha K_\bj \CD_\dalpha \bPhi^\bj
\end{align}
which can simplify its variation. We find after a lot of algebra
\begin{align*}
\delta \K_{\alpha \dalpha} =& -{H_{\alpha \dalpha}}^b \K_b
	-i V^\beta \CD_\beta \K_{\alpha \dalpha} + i V_\dbeta \CD^\dbeta \K_{\alpha \dalpha}
	- \Delta_{\alpha \dalpha} (V^b \K_b) \eol
	& + 2 \K_{i \bj} (V+\Sigma)\phi^i \CD_{\alpha \dalpha} \bphi^\bj
	 + 2 \K_{i \bj} (V+\Sigma)\bphi^\bj \CD_{\alpha \dalpha} \phi^i \eol
	& - \Delta_{\alpha \dalpha} (\Sigma^r \K_r) + (\Delta_{\alpha \dalpha} \Sigma^r)  \K_r \eol
	& - i \CD_\alpha {V_\dalpha}^\beta W_\beta^r \K_r
	+ i \CD_\dalpha {V_\alpha}^\dbeta W_\dbeta^r \K_r \eol
	& - \frac{i}{2} {V_{\dalpha}}^\beta (\CD_\alpha W_\beta^r) \K_r
	 + \frac{i}{2} {V_{\alpha}}^\dbeta (\CD_\dalpha W_\dbeta^r) \K_r \eol
	& - 2 \CD_\alpha {V_{\dalpha}}^\beta X_\beta^{(\K)}
	+ 2 \CD_\dalpha {V_{\alpha}}^\dbeta X_\dbeta^{(\K)} \eol
	& - {V_{\dalpha}}^\beta \CD_\alpha X_\beta^{(\K)}
	+ {V_{\alpha}}^\dbeta \CD_\dalpha X_\dbeta^{(\K)} \eol
	& + \CD_\alpha \phi^i \CD_\dalpha (\K_{i \bj} \bar\eta^\bj)
	- \CD_\dalpha \bphi^\bj \CD_\alpha (\K_{i \bj} \eta^i)
\end{align*}
where again we have collected a number of terms into the combination
$H_a{}^b$. The object $X_\beta^{(\K)}$ is defined as
\begin{align}
X_\beta^{(\K)} \equiv -\frac{1}{8} \bar\nabla^2 \nabla_\beta \K
\end{align}
In the chiral model, this can further be identified as the $U(1)$ spinor
field strength $X_\beta$.

Including the variation of the compensator and $V^a$ gives
\begin{align}
X^{-1} \delta (X V^a \K_a) =
	& + 2i V^\beta \CD_\beta V^a \K_a
	- 2i V_\dbeta \CD^\dbeta V^a \K_a
	+ 4 V^\alpha V^\dalpha \K_{\alpha \dalpha} \eol
	& + \tilde\delta U V^a \K_a
	- 2 V^a \K_a \Delta_b V^b \eol
	& + 2 \K_{i \bj} (V+\Sigma)\Phi^i V \bPhi^\bj
	 + 2 \K_{i \bj} (V+\Sigma)\bPhi^\bj V \Phi^i \eol
	& - \Delta_b V^b (\Sigma^r \K_r)
	+ V^a (\Delta_a \Sigma^r) \K_r \eol
	& - V^{\dalpha \alpha} \left(-\frac{1}{2} \CD_\alpha V^b F_{b \alpha} + \frac{1}{2} \CD_\dalpha V^b F_{b \dalpha}
	- \frac{1}{4} V^b \CD_\alpha F_{b \dalpha}
	+ \frac{1}{4} V^b \CD_\dalpha F_{b \alpha} \right) \K_r \eol
	& + V^{\dalpha \alpha} \CD_\alpha {V_{\dalpha}}^\beta X_\beta^{(\K)}
	- V^{\dalpha \alpha} \CD_\dalpha {V_{\alpha}}^\dbeta X_\dbeta^{(\K)} \eol
	& + \frac{1}{4} V^{\dalpha \alpha} V_{\alpha \dalpha}
	\left(\CD^\beta X_\beta^{(\K)} + \CD_\dbeta X^\dbeta{}^{(\K)}\right) \eol
	& - \frac{1}{2} V^{\dalpha \alpha} \CD_\alpha \Phi^i \CD_\dalpha (\K_{i \bj} \bar\eta^\bj)
	+ \frac{1}{2} V^{\dalpha \alpha} \CD_\dalpha \bPhi^\bj \CD_\alpha (\K_{i \bj} \eta^i)
\end{align}

The term arising from varying the Yang-Mills piece is fairly complicated.
One finds
\begin{align}
\delta \mathcal Y_{\alpha \dalpha} =& -{H_{\alpha \dalpha}}^b \mathcal Y_b - (\Delta_{\alpha \dalpha} V^b) \mathcal Y_b
	-i \CD^\beta (V_\beta \mathcal Y_{\alpha \dalpha}) + i \CD_\dbeta (V^\dbeta \mathcal Y_{\alpha \dalpha}) \eol
	& + \frac{1}{4} \CD^\phi \CD_{\{\dalpha} V_{\dbeta\} \phi} \mathcal Y^\dbeta_\alpha
	- \frac{1}{4} \CD^\dphi \CD_{\{\alpha} V_{\beta\} \dphi} \mathcal Y^\beta_\dalpha \eol
	& - V^{\dbeta \beta} G_{\alpha \dbeta} \mathcal Y_{\beta \dalpha}
	- V^{\dbeta \beta} G_{\beta \dalpha} \mathcal Y_{\alpha \dbeta} \eol
	& + \frac{1}{4} G_{rs} \bar\nabla^2 (V_{\alpha \dbeta} W^{\dbeta r}) W_\dalpha^s
	- \frac{1}{4} G_{rs} \nabla^2 (V_{\dalpha \beta} W^{\beta r}) W_\alpha^s \eol
	& + i V^b \CD_b (f_{rs} W_{\alpha}{}^r) W_\dalpha^s + i V^b \CD_b W_{\alpha}^r W_{\dalpha}{}^s \bar f_{rs}
	- i V^b W_{\alpha}{}^r f_{rs} \CD_b W_\dalpha^s - i V^b W_{\alpha}^r \CD_b (\bar f_{rs} W_{\dalpha}{}^s) \eol
	& - (G_{rs,i} \eta^i + G_{rs,\bj} \bar\eta^\bj) W_\alpha^r W_\dalpha^s \eol
	& - 2i f_{rs} W_\alpha^r \Sigma W_\dalpha^s
	+ 2 i \bar f_{rs} (\Sigma W_\alpha^r) W_\dalpha^s
	+ \frac{i}{4} G_{rs} \bar\nabla^2 \nabla_\alpha \Sigma^r W_\dalpha^s
	- \frac{i}{4} G_{rs} \nabla^2 \nabla_\dalpha \Sigma^r W_\alpha^s
\end{align}
A number of somewhat complicated looking terms have been introduced
in the first few lines, partly because the $H_a{}^b$ term is not
generated here as readily as in $\delta G_a$ and $\delta \K_a$.
A more convenient arrangement of the above terms is given by
\begin{align*}
\delta \mathcal Y_{\alpha \dalpha} =
	& -{H_{\alpha \dalpha}}^b \mathcal Y_b
	-i \CD^\beta (V_\beta \mathcal Y_{\alpha \dalpha}) + i \CD_\dbeta (V^\dbeta \mathcal Y_{\alpha \dalpha}) \\
	& -4 (\CD_b V_c) G_d^{\mathcal Y} \eps_{(\alpha \dalpha) b c d}
	+ \CD^\beta (G^{\mathcal Y}_{\alpha \dalpha} \CD^\dbeta V_{\alpha \dalpha})
	- \CD^\dbeta (G^{\mathcal Y}_{\alpha \dalpha} \CD^\beta V_{\alpha \dalpha}) \eol
	& + 2 \CD^\beta ( R^{\mathcal Y} \CD_\beta V_{\alpha \dalpha})
	+ 2 \CD_\dbeta ( \bar R^{\mathcal Y} \CD^\dbeta V_{\alpha \dalpha})
	- \CD^\gamma V^{\dbeta \beta} W^{\mathcal Y}_{\gamma (\beta \dbeta) (\alpha \dalpha)}
	- \CD^\dgamma V^{\dbeta \beta} W^{\mathcal Y}_{\dgamma (\beta \dbeta) (\alpha \dalpha)} \eol
	& + 2\left(-\frac{1}{2} \CD_\alpha V^b F_{b \alpha}{}^r + \frac{1}{2} \CD_\dalpha V^b F_{b \dalpha}{}^r
	- \frac{1}{4} V^b \CD_\alpha F_{b \dalpha}{}^r
	+ \frac{1}{4} V^b \CD_\dalpha F_{b \alpha}{}^r \right) \mathcal Y_r \eol
	& - 2 V^{\dbeta \beta} \mathcal Y_{(\alpha \dalpha) (\beta \dbeta)}
	- (G_{rs,i} \eta^i + G_{rs,\bj} \bar\eta^\bj) W_\alpha^r \bar W_\dalpha^s \\
	& - 2 i f_{rs} W_\alpha^r \Sigma \bar W_\dalpha^s
	+ 2i \bar f_{rs} (\Sigma W_\alpha^r) \bar W_\dalpha^s
	+ \frac{i}{4} G_{rs} \bar\nabla^2 \nabla_\alpha \Sigma^r \bar W_\dalpha^s
	- \frac{i}{4} G_{rs} \nabla^2 \nabla_\dalpha \Sigma^r W_\alpha^s
\end{align*}
where we have made a number of definitions. In particular,
\begin{gather}
R^{\mathcal Y} \equiv -\frac{1}{16} G_{rs} W^\phi{}^r W_\phi{}^s \\
\bar R^{\mathcal Y} \equiv -\frac{1}{16} G_{rs} \bar W_\dphi{}^r \bar W^\dphi{}^s \\
G^{\mathcal Y}_{\alpha \dalpha} \equiv \frac{1}{4} G_{rs} W_\alpha{}^r \bar W_\dalpha{}^s
\end{gather}
These definitions should not be taken more seriously than just
serving as convenient names. $R^{\mathcal Y}$, for example, is not chiral unless
the gauge couplings are trivial. We have simply identified these combinations
since they seem like they shall combine nicely with actual objects of those names
in the graviton propagator.\footnote{It is plausible, although we haven't explored this
possibility deeply yet, that if the linear compensator is coupled to
the Chern-Simons term for the gauge sector, then the superfields $R$ and
$G$ defined in terms of $L$ will pick up contributions of the above form
for the case $G_{rs} \propto \delta_{rs}$.}
In addition, we have written ``curvature'' terms which will also combine
with the similar term in $\Box_V$:
\begin{align}
W^{\mathcal Y}_{\gamma (\beta \dbeta) (\alpha \dalpha)}
	\equiv \frac{1}{4} \eps_{\dbeta \dalpha} \sum_{\beta \alpha}
	\biggl(&
	\eps_{\alpha \gamma} W_\beta{}^r \CD_\dphi (G_{rs} \bar W^{\dphi s})
	+ \eps_{\alpha \gamma} W_\beta{}^r \CD^\phi (G_{rs} W_{\phi}{}^s) \eol
	& - \eps_{\alpha \gamma} G_{rs} W_\beta{}^r (\CD W)^s
	+ G_{rs} W_\alpha{}^r \CD_\gamma W_\beta{}^s
	\biggr)
\end{align}
\begin{align}
W^{\mathcal Y}_{\dgamma (\beta \dbeta) (\alpha \dalpha)}
	\equiv -\frac{1}{4} \eps_{\beta \alpha} \sum_{\beta \alpha}
	\biggl(&
	\eps_{\dalpha \dgamma} \bar W_\dbeta{}^r \CD_\dphi (G_{rs} \bar W^{\dphi s})
	+ \eps_{\dalpha \dgamma} \bar W_\dbeta{}^r \CD^\phi (G_{rs} W_{\phi}{}^s) \eol
	& - \eps_{\dalpha \dgamma} G_{rs} \bar W_\dbeta{}^r (\CD W)^s
	- G_{rs} \bar W_\dalpha{}^r \CD_\dgamma \bar W_\dbeta{}^s
	\biggr)
\end{align}
as well as the ``potential'' term
\begin{align}
V^{\dbeta \beta} \mathcal Y_{(\alpha \dalpha) (\beta \dbeta)}
	\equiv & - V^b G_b \mathcal Y_{\alpha \dalpha} - G_{\alpha \dalpha} V^b \mathcal Y_b
	+ V_{\alpha \dalpha} G^b \mathcal Y_b \eol
	& + \frac{1}{4} V^{\dbeta \beta} \left(\CD_\beta W_{\alpha r} \CD_{\dbeta} W_{\dalpha}^r
								+ \CD_\beta W_{\alpha}^r \CD_{\dbeta} W_{\dalpha r}\right)
	- \frac{1}{2} V^b \Delta_b \mathcal Y_{\alpha \dalpha} \eol
	& - \frac{1}{8} V_{\alpha \dbeta} (\BCD^2 - 8 R) \bar W^{\dbeta r}\, G_{rs} W_{\dalpha}{}^s
	+ \frac{1}{8} V_{\dalpha \beta} (\CD^2 - 8 \bar R) W^{\beta r}\, G_{rs} W_{\alpha}{}^s \eol
	& + \frac{1}{8} V_{\dalpha \beta} \CD_\alpha W^{\beta r} \CD^\gamma (f_{rs} W_\gamma{}^s)
	+ \frac{1}{8} V_{\dalpha \beta} \CD_\alpha W^{\beta r} \CD_\dgamma (\bar f_{rs} \bar W^\dgamma{}^s) \eol
	& - \frac{1}{8} V_{\alpha \dbeta} \CD_\dalpha \bar W^{\dbeta r} \CD^\gamma (f_{rs} W_\gamma{}^s)
	- \frac{1}{8} V_{\alpha \dbeta} \CD_\dalpha \bar W^{\dbeta r} \CD_\dgamma (\bar f_{rs} \bar W^\dgamma{}^s)
\end{align}
These look like they could be defined in terms of the new $R^{\mathcal Y}$ and $G^{\mathcal Y}$
objects we have mentioned before, but we will avoid doing so explicitly.

The combination we need is
\begin{align}
\delta (V^a \mathcal Y_a) =
	& 2 i V^\beta \CD_\beta V^a \mathcal Y_a 
	- 2 i V_\dbeta \CD^\dbeta V^a \mathcal Y_a
	+ 4 V^\alpha V^\dalpha \mathcal Y_{\alpha \dalpha} \eol
	& -4 V^a (\CD_b V_c) G_d^{\mathcal Y} \eps_{a b c d} \eol
	& - \frac{1}{2} V^{\dalpha \alpha} \CD^\beta (G^{\mathcal Y}_{\beta \dbeta} \CD^\dbeta V_{\alpha \dalpha})
	+ \frac{1}{2} V^{\dalpha \alpha} \CD^\dbeta (G^{\mathcal Y}_{\alpha \dalpha} \CD^\beta V_{\alpha \dalpha}) \eol
	& - V^{\dalpha \alpha} \CD^\beta ( R^{\mathcal Y} \CD_\beta V_{\alpha \dalpha})
	 - V^{\dalpha \alpha} \CD_\dbeta ( \bar R^{\mathcal Y} \CD^\dbeta V_{\alpha \dalpha})\eol
	& + \frac{1}{2} V^{\dalpha \alpha} \CD^\gamma V^{\dbeta \beta} W^{\mathcal Y}_{\gamma (\beta \dbeta) (\alpha \dalpha)}
	+ \frac{1}{2} V^{\dalpha \alpha} \CD^\dgamma V^{\dbeta \beta} W^{\mathcal Y}_{\dgamma (\beta \dbeta) (\alpha \dalpha)} \eol
	& - V^{\dalpha \alpha} \left(-\frac{1}{2} \CD_\alpha V^b F_{b \alpha} + \frac{1}{2} \CD_\dalpha V^b F_{b \dalpha}
	- \frac{1}{4} V^b \CD_\alpha F_{b \dalpha}
	+ \frac{1}{4} V^b \CD_\dalpha F_{b \alpha} \right) \mathcal Y_r \eol
	& + V^{\dalpha \alpha} V^{\dbeta \beta} \mathcal Y_{(\alpha \dalpha) (\beta \dbeta)}
	+ \frac{1}{2} V^{\dalpha \alpha} (G_{rs,i} \eta^i + G_{rs,\bj} \bar\eta^\bj) W_\alpha^r \bar W_\dalpha^s \eol
	& + i V^{\dalpha \alpha} f_{rs} W_\alpha^r \Sigma \bar W_\dalpha^s
	- i V^{\dalpha \alpha} \bar f_{rs} (\Sigma W_\alpha^r) \bar W_\dalpha^s
	- \frac{i}{8} V^{\dalpha \alpha} G_{rs} \bar\nabla^2 \nabla_\alpha \Sigma^r \bar W_\dalpha^s
	+ \frac{i}{8} V^{\dalpha \alpha} G_{rs} \nabla^2 \nabla_\dalpha \Sigma^r W_\alpha^s
\end{align}

\subsubsection{Variation of the $\mathcal L$ term}
In the simple linear compensator model, there is one additional term -- that
involving $\mathcal L$. Beginning with
\begin{align}
S_L = \left[\mathcal L (V_R + K) \right]_D
\end{align}
one varies it to find
\begin{align}\label{eq_linterm}
\delta S_L = \mathcal L \left(
	3 \frac{\mathcal L}{L}
	- 2 \Delta_b V^b + V^b (K_b - 4 G_b)
	+ K_i \eta^i + K_\bj \bar\eta^\bj + \Sigma^r K_r
	\right)
\end{align}

\subsection{Summary}\label{summary}
We will break down our results into various sectors.

The terms involving just the chiral (and antichiral) quanta are
\begin{align*}
S_{\eta \eta} = 
	& \left[\eta^i X \K_{i \bar j} \bar\eta^\bj\right]_D
	+ \left[\eta^i (\CP (X \K_{i j}) + P_{ij} + \mathcal Y_{ij} ) \eta^j\right]_F + \hc
\end{align*}

The terms involving chiral and gauge fields are
\begin{align*}
S_{\eta \Sigma} &= 4 i \eta^i \K_{i \bj} X_r \bPhi^\bj \Sigma^r
	+ i \eta^i f_{rs,i} W^{\alpha s} \nabla_\alpha \Sigma^r + \hc \eol
	&= 2 \eta^i \left(X \K_{ir} + \mathcal Y_{ir}\right) \Sigma^r + \hc 
\end{align*}

The terms involving chiral and gravity fields are
\begin{align*}
S_{\eta V} &=
	+ V^{\dalpha \alpha} \CD_\dalpha \bPhi^\bj \CD_\alpha (X \K_{i \bar j} \eta^i)
	+ V^{\dalpha \alpha} W_{\alpha}^r \bar W_\dalpha^s f_{rs,i} \eta^i + \hc \eol
	&= 2 V^a \left(X \K_{a,i} \eta^i + \mathcal Y_{a,i} \right) \eta^i + \hc
\end{align*}

The terms involving gravity and gauge fields are
\begin{align}
S_{\Sigma V} = 
	& \left(2i V^\alpha \CD_\alpha \Sigma^r - 2i V_\dalpha \CD^\dalpha \Sigma^r\right)(X \K_r + \mathcal Y_r) \eol
	& - 2 V^a (\Delta_a \Sigma^r) \mathcal Y_r - 2X (\Delta_b V^b) \Sigma^r \K_r \eol
	& + \frac{i}{4} V_{\alpha \dalpha} G_{rs} \bar W^{\dalpha s} \bar\nabla^2 \nabla^\alpha \Sigma^r 
	- \frac{i}{4} V^{\dalpha \alpha} G_{rs} W_{\alpha}{}^s \nabla^2 \nabla_\dalpha \Sigma^r \eol
	& 
	+ 4 X \K_{i \bj} V \Phi^i \Sigma \bPhi^\bj
	+ 4 X \K_{i \bj} \Sigma \Phi^i V \bPhi^\bj  \eol
	& + 2i V^{\dalpha \alpha} f_{rs} W_\alpha^r \Sigma \bar W_\dalpha^s
	- 2i V^{\dalpha \alpha} \bar f_{rs} (\Sigma W_\alpha^r) \bar W_\dalpha^s
\end{align}
In the last two lines, we use a single $\Sigma$ to denote $\Sigma^r X_r$
acting to the right.
It seems reasonable to rearrange the second line of $S_{\Sigma V}$ so
that it is proportional to the equation of motion.
\begin{align}
S_{\Sigma V} = 
	& \left(2i V^\alpha \CD_\alpha \Sigma^r - 2i V_\dalpha \CD^\dalpha \Sigma^r\right)(X \K_r + \mathcal Y_r) \eol
	& - 2(\Delta_b V^b) \Sigma^r (X \K_r + \mathcal Y_r) \eol
	& + \frac{i}{4} V_{\alpha \dalpha} G_{rs} \bar W^{\dalpha s} \bar\nabla^2 \nabla^\alpha \Sigma^r 
	- \frac{i}{4} V^{\dalpha \alpha} G_{rs} W_{\alpha}{}^s \nabla^2 \nabla_\dalpha \Sigma^r \eol
	& - \Sigma^r \CD^\alpha V_{\alpha \dalpha} \CD^\dalpha \mathcal Y_r
	+ \Sigma^r \CD^\dalpha V_{\alpha \dalpha} \CD^\alpha \mathcal Y_r \eol
	& - 2 \Sigma^r V^a \Delta_a \mathcal Y_r \eol
	& 
	+ 4 X\K_{i \bj} V \Phi^i \Sigma \bPhi^\bj
	+ 4 X\K_{i \bj} \Sigma \Phi^i V \bPhi^\bj  \eol
	& + 2i V^{\dalpha \alpha} f_{rs} W_\alpha^r \Sigma \bar W_\dalpha^s
	- 2i V^{\dalpha \alpha} \bar f_{rs} (\Sigma W_\alpha^r) \bar W_\dalpha^s
\end{align}
The term with three spinor derivatives can be rearranged so that it is proportional to
$\CD^\alpha V_{\alpha \dalpha} (\BCD^2 - 8 R) \Sigma^r G_{rs} \bar W^\dalpha{}^s$, which can
be cancelled if we introduce a Gaussian smearing with the gauge fixing functions
$\CD^\alpha V_{\alpha \dalpha}$ for the gravity sector and
$(\BCD^2 - 8 R) \Sigma^r$ for the gauge sector, which is the standard
approach. \cite{superspace}

Next we turn to the pure gauge sector. We find
\begin{align}
S_{\Sigma \Sigma} =&
	4 X \K_{i \bj} \Sigma \Phi^i \Sigma \bPhi^\bj
	+ \Sigma^r \CD^\alpha \Sigma^s f_{sr}{}^t W_{\alpha}{}^t f_{tu}
	- \Sigma^r \CD_\dalpha \Sigma^s f_{sr}{}^t \bar W^{\dalpha}{}^t \bar f_{tu} \eol
	&
	- \frac{1}{8} \Sigma^r \nabla^\alpha (f_{rs} \bar\nabla^2 \nabla_\alpha \Sigma^s)
	- \frac{1}{8} \Sigma^r \nabla_\dalpha (\bar f_{rs} \nabla^2 \nabla^\dalpha \Sigma^s)
\end{align}
It is conspicous that for arbitrary holomorphic $f_{rs}$, the last term
yields the three spinor derivative term
$\Sigma^r (\nabla^\alpha f_{rs}) \bar\nabla^2 \nabla_\alpha \Sigma^s$
which it does not seem possible to remove by a smeared gauge. It is not
strictly speaking problematic to have a third order spinor derivative term
(as it is still less divergent than the pure kinetic term and so can
in principle be treated at least perturbatively), but it will lead to a more
complicated one-loop analysis.

In any case, it is useful to rearrange the kinetic term into a form
involving chiral projections of $\Sigma$. We use the identity
\begin{align} \label{eq_DV}
\frac{1}{8} \Sigma \nabla^\alpha (f \bar\nabla^2 \nabla_\alpha \Sigma) + \hc =& 
	(\CD_a \Sigma) G (\CD_a \Sigma)
	+ \frac{1}{8} (\BCD^2 - 8 R) \Sigma G (\CD^2 - 8 \bar R) \Sigma \eol
	& + \left(\frac{1}{8} \BCD_\dalpha \Sigma \BCD^\dalpha \bar f (\CD^2 - 8 \bar R) \Sigma
	+ \hc\right)
	 \eol
	& - 8 R \bar R \Sigma G \Sigma
	+ \frac{1}{2} \Sigma \CD^\alpha f \Sigma \CD_\alpha R
	+ \frac{1}{2} \Sigma \CD_\dalpha \bar f \Sigma \CD^\dalpha \bar R
	+ \frac{1}{2} \Sigma G \Sigma (\CD^2 R + \BCD^2 \bar R) \eol
	& - \CD_\alpha \Sigma G^{\dalpha \alpha} G \CD_\dalpha \Sigma
	+ \frac{i}{4} \CD^\alpha \Sigma \CD^\dalpha \Sigma
		(\CD_{\alpha \dalpha} f - \CD_{\alpha \dalpha} \bar f) \eol
	& + \CD^\alpha \Sigma \bar f (W_\alpha \Sigma)
	- \CD_\dalpha \Sigma f (\bar W^\dalpha \Sigma) \eol
	& + \frac{i}{4} \BCD^\dalpha \Sigma \CD^\alpha f \CD_{\alpha \dalpha} \Sigma
	+ \frac{i}{4} \CD_\alpha \Sigma \CD_\dalpha \bar f \CD^{\dalpha \alpha} \Sigma
\end{align}
In the above, we have suppressed all gauge indices for the sake of a less cluttered
notation. They should be contracted in the obvious way, taking care to note that
$(W_\alpha \Sigma)^r \equiv -W_\alpha^s \Sigma^t f_{ts}{}^r$. We have also
chosen to integrate certain terms by parts so that the result is manifestly
symmetric.

It is useful to define a generalized d'Alembertian for $\Sigma$
based on the above formula. We choose
\begin{align}\label{eq_boxV}
\Box^V_{rs} \Sigma^s \equiv &
	\CD^a (G_{rs} \CD_a \Sigma^s)
	- \frac{1}{2} \CD^{[\alpha} (G_{\alpha \dalpha} G_{rs} \CD^{\dalpha]} \Sigma^s)
	+ \CD^\alpha \Sigma^s G_{su} W_{\alpha}{}^t f_{r t}{}^u
	- \CD_\dalpha \Sigma^s G_{su} \bar W^\dalpha{}^t f_{r t}{}^u
\end{align}
so that in compacted notation
\begin{align}
\Sigma \Box_V \Sigma =
	\Sigma \CD^a (G \CD_a \Sigma)
	- \frac{1}{2} \Sigma \CD^{[\alpha} (G_{\alpha \dalpha} G \CD^{\dalpha]} \Sigma)
	- \CD^\alpha \Sigma G W_\alpha \Sigma
	+ \CD_\dalpha \Sigma G \bar W^\dalpha \Sigma
\end{align}
This is a generalization
of the scalar d'Alembertian $\Box_V$ discussed
in \cite{Buchbinder:1998qv}, generalized to a superfield $\Sigma$ with a nontrivial
gauge sector with corresponding gaugino superfield $W_\alpha$.
The form of this operator also inspired the definition of
$\Box_V V_{\alpha \dalpha}$ for the gravity sector.

We may then write
\begin{align*}
S_{\Sigma \Sigma} =&
	\, \Sigma \Box_V \Sigma
	- \frac{1}{8} (\BCD^2 - 8 R) \Sigma G (\CD^2 - 8 \bar R) \Sigma \eol
	& - \frac{1}{2} \Sigma G \Sigma (\CD^2 - 8 \bar R) R
	- \frac{1}{2} \Sigma G \Sigma (\BCD^2 - 8 R) \bar R
	+ 4 X \K_{i \bj} \Sigma \Phi^i \Sigma \bPhi^\bj \eol
	& - \left(\frac{1}{8} \BCD_\dalpha \Sigma \BCD^\dalpha \bar f (\CD^2 - 8 \bar R) \Sigma
	+ \hc\right)
	 \eol
	& - \frac{1}{2} \Sigma \CD^\alpha f \Sigma \CD_\alpha R
	- \frac{1}{2} \Sigma \CD_\dalpha \bar f \Sigma \CD^\dalpha \bar R
	- \frac{i}{4} \CD^\alpha \Sigma \CD^\dalpha \Sigma
		(\CD_{\alpha \dalpha} f - \CD_{\alpha \dalpha} \bar f) \eol
	& - \frac{i}{4} \BCD^\dalpha \Sigma \CD^\alpha f \CD_{\alpha \dalpha} \Sigma
	- \frac{i}{4} \CD_\alpha \Sigma \CD_\dalpha \bar f \CD^{\dalpha \alpha} \Sigma \eol
	& + \Sigma^r \CD^\alpha \Sigma^s W_\alpha{}^u (X_r f_{su})
	- \Sigma^r \CD_\dalpha \Sigma^s \bar W^\dalpha{}^u (X_r \bar f_{su})
\end{align*}
Note the last line involves the gauge generator acting on the holomorphic
gauge couplings. If these are taken to be proportional to the identity, then
the last line will vanish.

We turn finally to the pure gravity sector. The terms are quite numerous:
\begin{align}
S_{VV} =
	& \left(2i V^\alpha V^b F_{b \alpha}
	+ 2i V^\dalpha V^b F_{b \dalpha} \right) (X \K_r + \mathcal Y_r) \eol
	& \left(+ 2i V^\beta \CD_\beta V^a 
	- 2i V_\dbeta \CD^\dbeta V^a
	+ 4 V^\alpha V^\dalpha \sigma^a_{\alpha \dalpha} \right) \left(- 4 X G_a + X \K_a + \mathcal Y_a\right) \eol
	& + 2X (\Delta_b V^b)^2
	- 2X (\CD_b V^b)^2
	- \frac{X}{8} \CD^2 V^{\dalpha \alpha} \BCD^2 V_{\alpha \dalpha}
	+ X V^{\dalpha \alpha} \Box_V V_{\alpha \dalpha} \eol
	& - 2 \Delta_b V^b \left(X V^a \K_a - 4 X V^a G_a\right) \eol
	& + X V^{\dalpha \alpha} \CD^\beta (R \CD_\beta V_{\alpha \dalpha})
	+ X V^{\dalpha \alpha} \CD_\dbeta (\bar R \CD^\dbeta V_{\alpha \dalpha})  \eol
	& - V^{\dalpha \alpha} \CD^\beta ( R^{\mathcal Y} \CD_\beta V_{\alpha \dalpha})
	- V^{\dalpha \alpha} \CD_\dbeta ( \bar R^{\mathcal Y} \CD^\dbeta V_{\alpha \dalpha})\eol
	& - \frac{1}{2} V^{\dalpha \alpha} \CD^\beta (G^{\mathcal Y}_{\beta \dbeta} \CD^\dbeta V_{\alpha \dalpha})
	+ \frac{1}{2} V^{\dalpha \alpha} \CD^\dbeta (G^{\mathcal Y}_{\alpha \dalpha} \CD^\beta V_{\alpha \dalpha}) \eol
	& -4 V^a (\CD_b V_c) G_d^{\mathcal Y} \eps_{a b c d} \eol
	& + X V^{\dalpha \alpha} \CD_{(\alpha} {V_{\dalpha}}^\beta \hat\K_{\beta)}
	- X V^{\dalpha \alpha} \CD_{(\dalpha} {V_{\alpha}}^\dbeta \hat\K_{\dbeta)} \eol
	& + \frac{1}{2} V^{\dalpha \alpha} \CD^\gamma V^{\dbeta \beta} W^{\mathcal Y}_{\gamma (\beta \dbeta) (\alpha \dalpha)}
	+ \frac{1}{2} V^{\dalpha \alpha} \CD^\dgamma V^{\dbeta \beta} W^{\mathcal Y}_{\dgamma (\beta \dbeta) (\alpha \dalpha)} \eol
	&
	- 2 X R \bar R V^{\dalpha \alpha} V_{\alpha \dalpha}
	+ 4 X \K_{i \bj} V \Phi^i V \bPhi^\bj
	+ V^{\dalpha \alpha} V^{\dbeta \beta} \mathcal Y_{(\alpha \dalpha) (\beta \dbeta)}
\end{align}
We have defined
\begin{align}
\hat \K_\alpha = \piecett{-\frac{1}{8} \bar\nabla^2 \nabla_\alpha K - X_\alpha}{\textrm{   for the simple linear compensator model}}
	{0}{\textrm{   for the arbitrary chiral model}}
\end{align}

We have until now left the gauge for $V^\alpha$ unspecified.
Inspection of its appearance in all the terms shows that it is
always proportional to the equations of motion, so if we work
with the background on-shell then the gauge of $V^\alpha$
(at least to one-loop order) is physically irrelevant.
We will still choose the particular gauge $V^\alpha = 0$ for
definiteness.

The above represent the common features of the linear and chiral
models. They also each have a term involving $\tilde \delta U$:
\begin{align*}
S_{\delta U, *} = \tilde\delta U (X \K_i \eta^i + X \K_{\bj} \bar\eta^\bj + X \K_r \Sigma^r
					- 4 X V^b G_b + X V^b \K_b - 2 X \Delta_b V^b)
\end{align*}
Depending on the model, the variation of the compensator may
be quite different. The simple linear compensator model has
\begin{align*}
\tilde \delta U = L^{-1} \mathcal L
\end{align*}
while the arbitrary chiral model possesses
\begin{align*}
\tilde \delta U = -\frac{1}{3} \left(\K_i \eta^i + \K_{\bj} \bar\eta^\bj + \K_r \Sigma^r
					- 4 V^b G_b + V^b \K_b - 2 \Delta_b V^b \right)
\end{align*}
In addition, for the linear compensator model there are the terms
arising from varying \eqref{eq_linterm}:
\begin{align}
S_{L,*} = \mathcal L \left(
	3 \frac{\mathcal L}{L}
	- 2 \Delta_b V^b + V^b (K_b - 4 G_b)
	+ K_i \eta^i + K_\bj \bar\eta^\bj + \Sigma^r K_r
	\right)
\end{align}

Combining these two effects gives the second order action for the
linear compensator model
\begin{align}
S_{\mathrm{linear}}^{(2)} =
	& \, S_{VV} + S_{\Sigma V} + S_{\Sigma \Sigma}
	+ S_{\eta V} + S_{\eta\Sigma} + S_{\eta \eta} \eol
	& + 3 \frac{\mathcal L^2}{L}
	+ 2 \mathcal L \left(
	K_i \eta^i + K_\bj \bar\eta^\bj + \Sigma^r K_r
	+ V^b (K_b - 4 G_b) - 2 \Delta_b V^b 
	\right)
\end{align}

For the chiral model, we find
\begin{align}
S_{\mathrm{chiral}}^{(2)} =
	& \,S_{VV} + S_{\Sigma V} + S_{\Sigma \Sigma}
	+ S_{\eta V} + S_{\eta\Sigma} + S_{\eta \eta}\eol
	& -\frac{X}{3} \left(\K_i \eta^i + \K_{\bj} \bar\eta^\bj + \K_r \Sigma^r
					+ V^b (\K_b - 4 G_b) - 2 \Delta_b V^b \right)^2
\end{align}

For reference, we include here their first order variations as well:
\begin{align}
S_{\mathrm{chiral}}^{(1)} =&\,
	\left[V^a (X \K_a - X G_a + \mathcal Y_a) + \Sigma^r (X \K_r + \mathcal Y_r) \right]_D \eol
	& + \left[\eta^i (\CP \K_i + P_i + \mathcal Y_i) \right]_F
	+ \left[\bar\eta^\bj (\ACP \K_\bj + \bar P_\bj + \mathcal Y_\bj) \right]_F \\
S_{\mathrm{linear}}^{(1)} =&\,
	\left[V^a (X K_a - X G_a + \mathcal Y_a) + \Sigma^r (X K_r + \mathcal Y_r) \right]_D \eol
	& + \left[\eta^i (\CP K_i + P_i + \mathcal Y_i) \right]_F
	+ \left[\bar\eta^\bj (\ACP K_\bj + \bar P_\bj + \mathcal Y_\bj) \right]_F \eol
	& + \left[\mathcal L (V_R + K) \right]_D
\end{align}

Their respective actions to second order in the quantum fields are
then given by
\begin{align}
S_{\mathrm{chiral}} = S_{\mathrm{chiral}}^{(0)} + S_{\mathrm{chiral}}^{(1)} + \frac{1}{2} S_{\mathrm{chiral}}^{(2)} \\
S_{\mathrm{linear}} = S_{\mathrm{linear}}^{(0)} + S_{\mathrm{linear}}^{(1)} + \frac{1}{2} S_{\mathrm{linear}}^{(2)}
\end{align}

When we consider that the linear compensator model is classically
dual to a special case of the arbirary chiral model, it becomes
perhaps unsurprising that their quantum actions should have so many
features in common. This commonality is enough for us to ask whether
the two theories might actually be equivalent at the one-loop
level, at least on-shell. One can in fact make a rather straightforward
argument, based on the existing proofs of equivalence for
chiral spinors and chiral scalars \cite{Buchbinder:1998qv, Buchbinder:1988tj}
that the two effective actions should be equivalent on-shell at one-loop.
There is a subtlety, however, due to the inability to nicely
define the path integration for a generic chiral superfield. 
We will return to this issue in a subsequent paper.

\section{Conclusion}
The formulae listed above constitute the end
of the algebraic manipulations necessary to produce a suitable
action quadratic in the quantum superfields of supergravity,
super Yang-Mills, and chiral matter. Further steps
are necessary to produce one-loop results.

The first step is obviously to perform a gauge-fixing of the gravity and gauge
sectors. Part of the procedure here will involve deciding just
\emph{how} to do it. Even if we choose a smeared gauge and
aim for only $1/p^2$ propagators (as was the guiding principle in
\cite{Grisaru:1981xm}), we have the option of removing 
certain terms in $S_{\Sigma V}$ or $S_{VV}$ involving operators of
dimension less than two. Any choice must, of course, be physically
equivalent to any other, but certain calculational simplifications
may occur only one way. It is possible that the dual formulation with
the linear compensator can play a role in helping us find the simplest
gauge choice due to the way in which it decouples the matter and
gravitational sectors, but we have not finished exploring other options
yet.

The second is to actually perform the resulting path integrals.
For background field calculations, one generally prefers a method
which is non-perturbative, such as the Schwinger proper time method
or the derivative expansion. Such a procedure here is a bit more
difficult since while the gauge and gravity sectors involve generalized
Laplacians, the chiral sector involves Dirac-like operators.
If the couplings between these sectors do not vanish, some
amount of perturbation seems necessary, since the determinant
of an operator with a diagonal consisting of Laplace \emph{and}
Dirac operators is difficult to deal with without separating
out the two sectors. We hope to explore these two steps soon.


\newpage
\appendix
\section{Arbitrary linear and chiral superfield models at first order}\label{arblin}
We have expanded the actions for arbitrary chiral models to
second order in the quantum fields to enable quantization.
The structure they possess is fairly interesting and is
reflected in the minimal model of a linear compensator
coupled to supergravity and a K\"ahler potential. We will
briefly consider the generalization to an \emph{arbitrary}
coupling of a linear superfield $L$ to chiral multiplets $\Phi^i$
in the context of conformal supergravity. Although we will assume
only a single linear superfield $L$, the generalization to several
is straightforward.

The interesting part will be contained in the D-term action
\begin{align*}
S_D = -3 \int E \, \mathcal F(L, \Phi^i, \bPhi^\bj)
\end{align*}
The $-3$ is chosen so that if $\mathcal F$ is independent of $L$, a
canonical Einstein-Hilbert term is reproduced for the choice $\mathcal F=1$.
Observing that
\begin{gather*}
D \mathcal F = 2 \mathcal F = \mathcal F_i \Delta_i \Phi^i
	+ \mathcal F_{\bj} \Delta_{\bj} \bPhi^\bj + 2 \mathcal F_L L \\
-\frac{3i}{2} A \mathcal F = 0 = \mathcal F_i \Delta_i \Phi^i
	- \mathcal F_{\bj} \Delta_{\bj} \bPhi^\bj 
\end{gather*}
and that the Einstein-Hilbert term is contained within
\[
S_D \ni -\frac{3}{2} \mathcal F_i \Box \Phi^i - \frac{3}{2} \mathcal F_{\bj} \Box \bPhi^\bj
\]
where $\Box$ are superconformal (and thus contain $\mathcal R/6$ weighted by the
scaling dimension of the field on which $\Box$ acts), it is easy
to see that the normalization of the Einstein-Hilbert term is
\[
X = \frac{1}{2} \mathcal F_i \Delta_i \Phi_i + \frac{1}{2} \mathcal F_\bj \Delta_{\bj} \bPhi^\bj
	= \mathcal F - L \mathcal F_L
\]
It is clear that the field multiplying the Einstein-Hilbert
term is the proper conformal compensator to use for our theory,
so we have chosen to label the above combination as $X$.

Expanding $S_D$ to first order in quantum fields using the tools
we have developed is straightforward. One finds
\begin{align*}
\delta S_D =& 3i \nabla^\alpha (V_\alpha \F) - 3i \nabla_\dalpha (V^\dalpha \F)
	+ 3\Delta_b (V^b L) \F_L + (\Delta_b V^b) (\F - L \F_L) \eol
	& + 3i V^b \left(\F_i \nabla_b \Phi^i - \F_\bj \nabla_b \bPhi^\bj\right)
	+ 3i \Sigma^r \left(\F_i X_r \Phi^i - \F_\bj X_r \bPhi^\bj\right)
	-3 \F_L \mathcal L
\end{align*}
where $\Delta_b$ is conformally covariant, as are all the other derivatives.
Integrating by parts (and taking care that the special conformal connections vanish)
gives
\begin{align*}
\delta S_D =&
	+ 3 V^b L \Delta_b \F_L + V^b \Delta_b (\F - L \F_L) 
	+ 3 i V^b \left(\F_i \nabla_b \Phi^i - \F_\bj \nabla_b \bPhi^\bj\right) \eol
	& + 3 i \Sigma^r \left(\F_i X_r \Phi^i - \F_\bj X_r \bPhi^\bj\right)
	- 3 \F_L \mathcal L
\end{align*}
Using
\begin{align*}
V^b \Delta_b \F =& V^b \F_L \Delta_b L - i V^b \F_i \nabla_b \Phi^i + i V^b \F_\bj \nabla_b \bPhi^\bj \eol
	& + \frac{1}{2} \mathcal F_{I\bar J} \nabla_\alpha \Psi^I \nabla_\dalpha \Psi^{\bar J}
\end{align*}
where $\Psi^I$ denotes the set $(\Phi^i, L)$ and $\Psi^{\bar J}$ the set $(\bPhi^\bj, L)$,
we can write the variation as
\begin{align*}
\delta S_D =&
	- 2 V^b \Delta_b (\F - L \F_L)
	+ \frac{3}{2} V^{\dalpha \alpha} \F_{i \bj} \nabla_\alpha \Phi^i \nabla_\dalpha \bPhi^\bj
	- \frac{3}{2} V^{\dalpha \alpha} \F_{L L} \nabla_\alpha L \nabla_\dalpha L \eol
	& + 3 i \Sigma^r \left(\F_i X_r \Phi^i - \F_\bj X_r \bPhi^\bj\right)
	- 3 \F_L \mathcal L
\end{align*}
This form is immediately reminiscent of that we have discussed before.
Since $X \equiv \F - L \F_L$ is to be identified as the compensator, we define
$G_a \equiv -X^{1/2} \Delta_a X^{-1/2}$ as before. This immediately yields
\begin{align*}
\delta S_D =&
	- 4 X V^b G_b
	+ \frac{3}{2} V^{\dalpha \alpha} \F_{i \bj} \nabla_\alpha \Phi^i \nabla_\dalpha \bPhi^\bj
	- \frac{3}{2} V^{\dalpha \alpha} \F_{L L} \nabla_\alpha L \nabla_\dalpha L
	- \frac{3}{2} V^{\dalpha \alpha} X^{-1} \nabla_\alpha X \nabla_\dalpha X \eol
	& + 3 i \Sigma^r \left(\F_i X_r \Phi^i - \F_\bj X_r \bPhi^\bj\right)
	- 3 \F_L \mathcal L
\end{align*}
To maintain the analogy, we should make the identifications
\begin{gather*}
K_{\alpha \dalpha} \equiv
	- 3 X^{-1} \F_{i \bj} \nabla_\alpha \Phi^i \nabla_\dalpha \bPhi^\bj
	+ 3 X^{-1} \F_{L L} \nabla_\alpha L \nabla_\dalpha L
	+ 3 X^{-2} \nabla_\alpha X \nabla_\dalpha X \eol
K_r \equiv + 3 i X^{-1} \left(\F_i X_r \Phi^i - \F_\bj X_r \bPhi^\bj\right)
\end{gather*}
which would give
\begin{align*}
\delta S_D =&
	- 4 X V^b G_b + X V^b K_b
	+ X \Sigma^r K_r
	- 3 \F_L \mathcal L
\end{align*}

We would like to think of terms involving $V^a$ to consist of
a ``supergravity term'' $G_b$ and the ``matter term'' $K_b$,
so it is sensible to expand $K_b$ out entirely in terms of
the fields. We find
\begin{align*}
K_{\alpha \dalpha} =&
	- 3 \nabla_\alpha \Phi^i \nabla_\dalpha \bPhi^\bj
			\left(X^{-1} \F_{i \bj} - X^{-2} X_i X_\bj \right) \eol
	& + 3 \nabla_\alpha \Phi^i \nabla_\dalpha L
			\left(X^{-2} X_i X_L \right)
	+ 3 \nabla_\alpha L \nabla_\dalpha \Phi^\bj
			\left(X^{-2} X_L X_\bj \right) \eol
	& + 3 \nabla_\alpha L \nabla_\dalpha L
			\left(X^{-1} \F_{L L} + X^{-2} X_L X_L\right)
\end{align*}
where $X_i = \mathcal F_i - L \mathcal F_{L i}$, 
$X_\bj = \mathcal F_\bj - L \mathcal F_{L \bj}$ and $X_L = -L \F_{LL}$.

Before moving on, we should make one more generalization. Up until now
we have assumed $L$ to be a normal linear multiplet. However, we may
instead choose for $L$ to obey the modified linearity condition
\begin{align}\label{eq_modlin}
\CP L = -\frac{1}{4} \bar\nabla^2 L = -\frac{1}{2} k \Tr(W^\alpha W_\alpha)
\end{align}
This amounts to choosing $L = L_0 + k \Omega$, where $L_0$ is a
normal linear superfield and $\Omega$ is the Chern-Simons superfield \cite{bgg}.
$L$ is chosen to be gauge invariant, so the gauge transformation of
$\Omega$, which is itself a linear superfield, must be cancelled by the
transformation of $L_0$.

The Yang-Mills term then receives contributions from the D-term of
$\mathcal F$:
\begin{align}
-3 \int E \, \mathcal F = \frac{3k }{4} \int \chE \, \biggl( \mathcal F_L \Tr(W^\alpha W_\alpha) + \ldots\biggr) + \hc
\end{align}
This contributes to $f_{rs}$ (effectively) a non-holomorphic factor of $3k\delta_{rs} \mathcal F_L$
and thus to $G_{rs}$ a factor of $6 k \delta_{rs} \mathcal F_L$.

The quanta of $L$ which we previously denoted $\mathcal L$ should now be understood as
\[
\mathcal L = \mathcal L_0 - i k \nabla^\alpha (W_\alpha\Sigma)
	- i k \nabla_\dalpha (W^\alpha \Sigma)
	+ i k (\nabla^\alpha W_\alpha) \Sigma - k V^{\dalpha \alpha} W_\alpha \bar W_\dalpha
\]
where $\mathcal L_0$ is linear. This formula is determined by requiring
the chiral quantum variation of both sides of \eqref{eq_modlin} to coincide.

One can easily check that
\begin{align*}
-3 \mathcal L \F_L &= -3 \F_L \mathcal L_0
	+ 3 i k \F_L \nabla^\alpha (W_\alpha\Sigma)
	+ 3 i k \F_L \nabla_\dalpha (W^\alpha \Sigma)
	- 3i k \F_L (\nabla^\alpha W_\alpha) \Sigma
	+ 3 k \F_L V^{\dalpha \alpha} W_\alpha \bar W_\dalpha \eol
	&=  -3 \F_L \mathcal L_0
	+ \Sigma^r \mathcal Y_r
	+ V^b \mathcal Y_b
\end{align*}
where
\begin{gather*}
\mathcal Y_r = -3i k (\nabla^\alpha \F_L) W_{\alpha r}
	- 3i k (\nabla^\dalpha \F_L) \bar W^\dalpha{}_r
	- 3i k \F_L (\nabla^\alpha W_\alpha){}_r \\
\mathcal Y_{\alpha \dalpha} = -6 \F_L k \Tr(W_\alpha \bar W_\dalpha)
\end{gather*}
This agrees with the previous definition for these objects provided
we rewrite them solely in terms of $G_{rs} = f_{rs} + \bar f_{rs}$
Then taking into account the contribution from the linear multiplet
gives $G'_{rs} = f_{rs} + \bar f_{rs} + 6 \F_L k \delta_{rs}$.

The first order structure can then be written
\begin{align}
\delta S_D =
	V^b \left(- 4 X G_b + X K_b + \mathcal Y_b\right)
	+ \Sigma^r \left( X K_r + \mathcal Y_r\right)
	- 3 \F_L \mathcal L
	- 3 \F_i \eta^i
	- 3 \F_\bj \bar\eta^\bj
\end{align}
where we have included also the chiral superfield variations.


\section*{Acknowledgments}
I am grateful to Mary K. Gaillard for helpful comments and discussions. This work was supported in part by the Director, Office of Science, Office of High Energy and Nuclear Physics, Division of High Energy Physics of the U.S. Department of Energy under Contract DE-AC02-05CH11231, in part by the National Science Foundation under grant PHY-0457315.


\begin{thebibliography}{99}

\bibitem{Grisaru:1981xm}
  M.~T.~Grisaru and W.~Siegel,
  ``Supergraphity. Part 1. Background Field Formalism,''
  Nucl.\ Phys.\  B {\bf 187}, 149 (1981).

\bibitem{Grisaru:1982zh}
  M.~T.~Grisaru and W.~Siegel,
  ``Supergraphity. 2. Manifestly Covariant Rules And Higher Loop Finiteness,''
  Nucl.\ Phys.\  B {\bf 201}, 292 (1982)
  [Erratum-ibid.\  B {\bf 206}, 496 (1982)].

\bibitem{Grisaru:1983rg}
  M.~T.~Grisaru and D.~Zanon,
  ``Quantum Superfield Supergravity With Off-Shell Background Fields,''
  Nucl.\ Phys.\  B {\bf 237}, 32 (1984).

\bibitem{Grisaru:1984ja}
  M.~T.~Grisaru and D.~Zanon,
  ``Covariant Supergraphs. 1. Yang-Mills Theory,''
  Nucl.\ Phys.\  B {\bf 252}, 578 (1985).

\bibitem{Grisaru:1984jc}
  M.~T.~Grisaru and D.~Zanon,
  ``Covariant Supergraphs. 2. Supergravity,''
  Nucl.\ Phys.\  B {\bf 252}, 591 (1985).


\bibitem{nkghost}
  N.~K.~Nielsen,
  ``Ghost Counting In Supergravity,''
  Nucl.\ Phys.\  B {\bf 140}, 499 (1978). \\
  R.~E.~Kallosh,
  ``Modified Feynman Rules In Supergravity,''
  Nucl.\ Phys.\  B {\bf 141}, 141 (1978).


\bibitem{hidden_ghosts}
  W.~Siegel,
  ``Hidden Ghosts,''
  Phys.\ Lett.\  B {\bf 93}, 170 (1980).

\bibitem{superspace}
  S.~J.~Gates, M.~T.~Grisaru, M.~Rocek and W.~Siegel,
  ``Superspace, or one thousand and one lessons in supersymmetry,''
  Front.\ Phys.\  {\bf 58}, 1 (1983)
  [arXiv:hep-th/0108200].

\bibitem{wb}
  J.~Wess and J.~Bagger,
  ``Supersymmetry and supergravity,''
{\it  Princeton, USA: Univ. Pr. (1992) 259 p}

\bibitem{bgg}
  P.~Binetruy, G.~Girardi and R.~Grimm,
  ``Supergravity couplings: A geometric formulation,''
  Phys.\ Rept.\  {\bf 343}, 255 (2001)
  [arXiv:hep-th/0005225].

\bibitem{Butter:2009cp}
  D.~Butter,
  ``N=1 Conformal Superspace in Four Dimensions,''
  arXiv:0906.4399 [hep-th].


\bibitem{Buchbinder:1998qv}
  I.~L.~Buchbinder and S.~M.~Kuzenko,
  ``Ideas and methods of supersymmetry and supergravity: Or a walk through
  superspace,''
{\it  Bristol, UK: IOP (1998) 656 p}


\bibitem{Ogievetsky:1978mt}
  V.~Ogievetsky and E.~Sokatchev,
  ``Structure of supergravity group,''
  Phys.\ Lett.\  {\bf 79B}, 222 (1978)
  [Czech.\ J.\ Phys.\  B {\bf 29}, 68 (1979)].


\bibitem{Kugo:1983mv}
  T.~Kugo and S.~Uehara,
  ``N=1 Superconformal Tensor Calculus: Multiplets With External Lorentz
  Indices And Spinor Derivative Operators,''
  Prog.\ Theor.\ Phys.\  {\bf 73}, 235 (1985).


\bibitem{Buchbinder:1988tj}
  I.~L.~Buchbinder and S.~M.~Kuzenko,
  ``Quantization Of The Classically Equivalent Theories In The Superspace Of
  Simple Supergravity And Quantum Equivalence,''
  Nucl.\ Phys.\  B {\bf 308}, 162 (1988).











\end{thebibliography}
\end{document}